# DIRECT MULTIPIXEL IMAGING AND SPECTROSCOPY OF AN EXOPLANET WITH A SOLAR GRAVITY LENS MISSION

NASA Innovative Advanced Concepts (NIAC) Phase I

## FINAL REPORT

Slava G. Turyshev[1], Michael Shao[1], Leon Alkalai[1], Nitin Aurora[1], Darren Garber[3], Henry Helvajian[2], Tom Heinsheimer[2], Siegfried Janson[2], Jared R. Males[6], Dmitri Mawet[1], Roy Nakagawa[2], Seth Redfield[5], Janice Shen[1], Nathan Strange[1], Mark R. Swain[1], Viktor T. Toth[4], Phil A. Willems[1], John L. West[1], Stacy Weinstein-Weiss[1], Hanying Zhou[1]

*[1]Jet Propulsion Laboratory, California Institute of Technology, 4800 Oak Grove Drive, Pasadena, CA 91109-0899, USA*

*[2]The Aerospace Corporation, El Segundo, CA*

*[3]NXTRAC Inc., Redondo Beach, CA 90277, https://nxtrac.com/*

*[4]Ottawa, Ontario K1N 9H5, Canada, www.vttoth.com*

*[5]Wesleyan University, 45 Wyllys Ave, Middletown, CT 06459, USA*

*[6]Department of Astronomy & Stewart Observatory, University of Arizona, Tucson AZ 8572, USA*

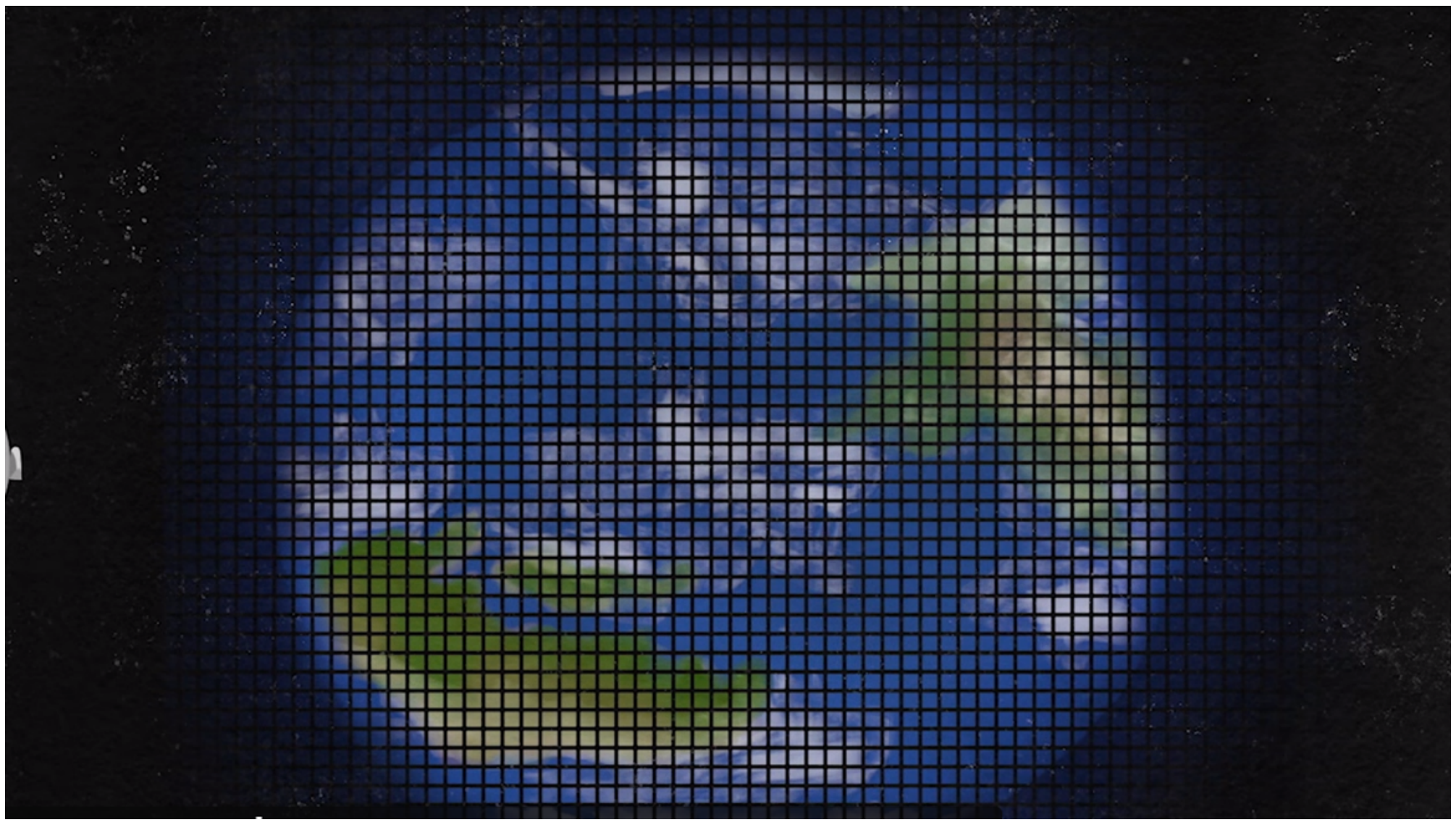

A one-meter telescope, with a modest coronagraph to block solar light with $10^{-6}$ suppression placed in the focal area of the solar gravitational lens, is capable to image an exoplanet at a distance up to 100 light years with kilometer-scale resolution on its surface. Notably, spectroscopic broadband SNR is $\sim 10^6$ in two weeks of integration time, providing this instrument with incredible remote sensing capabilities.

See concept video-description at [https://youtu.be/Hjaj-Ig9jBs](https://youtu.be/Hjaj-Ig9jBs)

**Executive Summary:**

The remarkable optical properties of the solar gravitational lens (SGL) include major brightness amplification (~$10^{11}$ at $\lambda$=1 μm) and extreme angular resolution (~$10^{-10}$ arcsec) in a narrow field of view (FOV). Such an instrument could benefit many areas of astrophysics involving exoplanets, star formation, nebula, accretion disks, neutron stars, galactic center, etc. Here we focus only on remote investigations of exoplanets.

A mission to the SGL carrying a modest telescope and coronagraph opens up a possibility for *direct megapixel* imaging and high-resolution spectroscopy of a habitable Earth-like exoplanet at a distance of up to 100 light years. The entire image of such a planet is compressed by the SGL into a region with a diameter of ~1.3 km in the vicinity of the focal line. The telescope, acting as a single pixel detector while traversing this region, can build an image of the exoplanet with kilometer-scale resolution of its surface, enough to see its surface features and signs of habitability. Although theoretically feasible, the engineering aspects of building and operating such an astronomical telescope on the large scales involved were not previously addressed. This unique concept requires serious consideration.

We report here on the results of our initial study of a mission to the deep outer regions of our solar system, with the primary mission objective of conducting direct megapixel high-resolution imaging and spectroscopy of a potentially habitable exoplanet by exploiting the remarkable optical properties of the SGL. Our main goal was not to study how to get there (although this was also addressed), but rather, to investigate what it takes to operate spacecraft at such enormous distances with the needed precision. Specifically, we studied i) how a space mission to the focal region of the SGL may be used to obtain high-resolution direct imaging and spectroscopy of an exoplanet by detecting, tracking, and studying the Einstein ring around the Sun, and ii) how such information could be used to detect signs of life on another planet.

We considered several mission concepts involving either i) a single probe class spacecraft, ii) a "string-of-pearls" mission concept using multiple sets of moderate size spacecraft, and iii) a swarm of small and capable spacecraft. Our results indicate that a mission to the SGL with an objective of direct imaging and spectroscopy of a distant exoplanet is challenging, but possible. We composed a list of recommendations on the mission architectures with risk and return tradeoffs and discuss an enabling technology development program.

Under a Phase I NIAC program, we evaluated the feasibility of the SGL-based technique for direct imaging and spectroscopy of an exoplanet and, while several practical constraints have been identified, we have not identified any fundamental limitations. We determined that the foundational technology already exists and has high TRL in space missions and applications. Furthermore, the measurements required to demonstrate the feasibility of remote sensing with the SGL are complementary to rotational tomography measurements and ongoing microlensing investigations, so our effort would provide high-value scientific information to other active astrophysics programs.

Our results are encouraging as they lead to a realistic design for a mission that will be able to image exoplanets. It could allow exploration of exoplanets relying on the SGL capabilities decades, if not centuries, earlier than possible with other extant technologies. The architecture and mission concepts for a mission to the SGL, at this early stage, are promising and should be explored further.

# 1 INTRODUCTION

We are standing at the threshold of a major discovery: The age-old question, "are we alone in the Universe?", may be answered within our lifetimes. Extensive evidence indicates that planets capable of harboring life are ubiquitous in our galaxy and are a standard phenomenon of a typical stellar evolution. Multiple discoveries of exoplanets orbiting nearby stars, efforts to understand the conditions that trigger, stimulate, and guide planetary formation processes, ignite their atmospheres and life-promoting conditions, as well as the development of techniques needed to find and study the new planets are hot topics at the focus of multiple ongoing science efforts. These efforts intensified with the recent discovery of the Earth-like planet Proxima Centauri b orbiting our stellar neighbor, triggering new activities to find intelligent life in the Universe. More such discoveries are to come from a multitude of ongoing efforts and efforts in development.

We are poised for more exciting news that may even include finding a habitable planet within the next decade. With the fleet of missions that are currently in space (Kepler), those that are being developed by NASA (TESS, 2017; JWST, 2018) and ESA (Euclid) and those that are being studied by the science community worldwide (HubEX, 2017; LUVOIR, 2017), the question now is not if, but when we will discover the signs of extraterrestrial life.

Direct detection, from the Earth or near-Earth space, of light reflected off the surface of a small and distant object moving in close proximity to its parent star is a very difficult task, which requires the development of dedicated technologies. The signal of interest is orders of magnitude weaker than that of its parent star, requiring coronagraphic techniques to eliminate the parasitic starlight. Because of the large distances involved, the planet subtends a very small angle in the sky, necessitating very large apertures to reach the required detection sensitivity. The light received from an exoplanet is extremely faint. Analyzing it requires detailed knowledge and a careful account of the relevant background noise which results in excessively long integration times to identify the signal. These technical challenges make direct high-resolution imaging of an exoplanet with conventional technologies—a telescope and an interferometer—a very difficult, if not impossible task.

Fortunately, nature has presented us with a powerful instrument that we can explore and learn to use for imaging purposes. This instrument is the solar gravitational lens (SGL), which takes advantage of the ability of the solar gravitational field to focus and greatly amplify light from sources of significant science interest, such as a possibly habitable exoplanet.

According to Einstein's general theory of relativity (Einstein, 1915; Einstein, 1916), the gravitational field induces refractive properties on spacetime (Fock, 1959), with a massive object acting as a lens, capable of bending the trajectories of incident photons (Turyshev, 2008). Experimental confirmation of the general relativistic gravitational bending of light trajectories nearly a century ago (Eddington, 1919; Dyson et al., 1920) unambiguously established that celestial bodies act as gravitational lenses focusing light from distant sources. The properties of gravitational lenses, including light amplification and the appearance of ring-like images (Einstein rings), are well established (Chwolson, 1924; Einstein, 1936; Herlt & Stephani, 1976, 1978) and have a rich literature (see (Turyshev & Toth, 2017) for review).

Today, gravitational lensing is a well-known and well-understood effect and has been observed over cosmological distances, where relatively nearby galaxies, or even clusters of galaxies, act as gravitational lenses for the light emitted by background galaxies. It has also been seen in our galaxy, where microlensing of stars in the galactic bulge or in the Magellanic clouds is caused by intervening stellar and substellar bodies. In the solar system, the effect was originally observed by Eddington in 1919. It is routinely accounted for in astronomy, astrophysics and spacecraft navigation. Astrometric microlensing is used to determine the masses of stellar objects. The time has arrived to start using it for the purposes of practical astronomy.

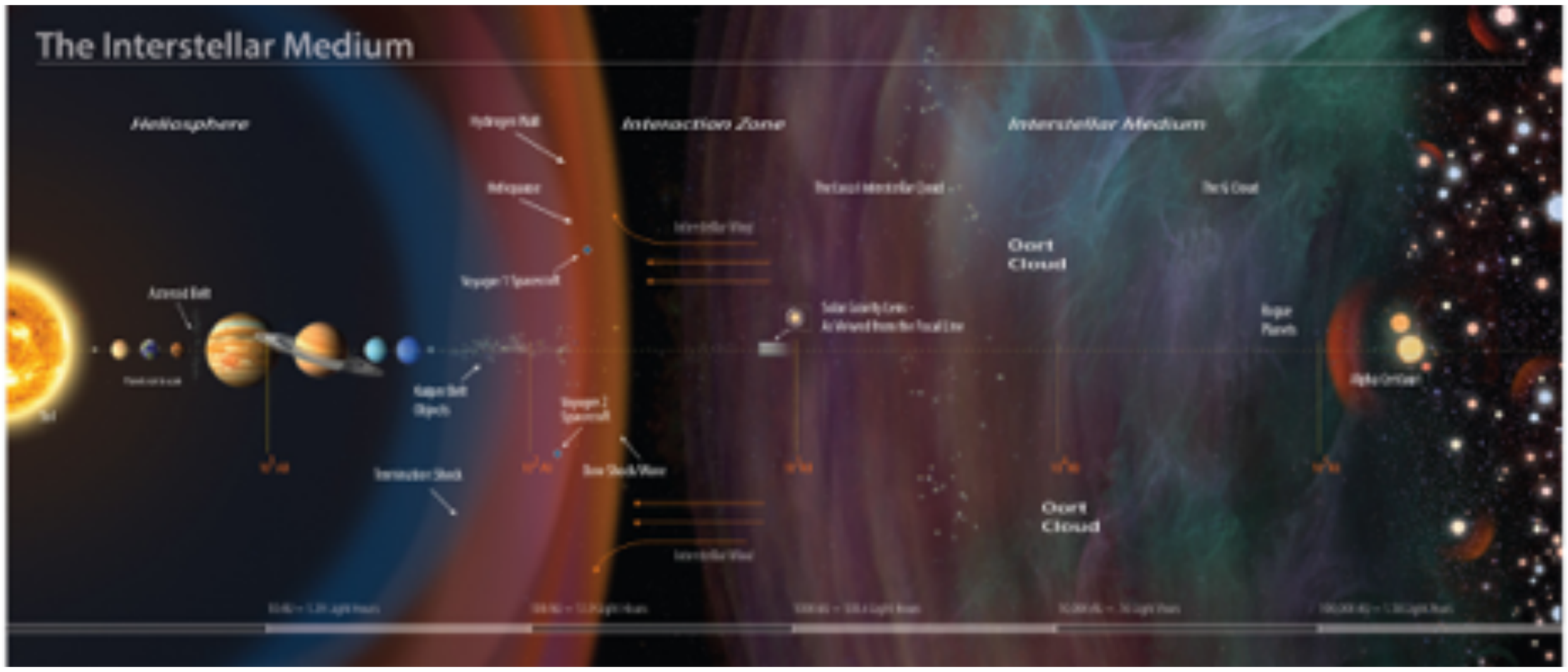

Figure 1. Our stellar neighborhood (in logarithmic scale) (Strone et al., 2015). The SGL is in the middle of this figure.

shows our stellar neighborhood including our solar system and going all the way to Alpha Centauri (Stone et al., 2015). The SGL is in the middle of this picture.

In the geometrical optics approximation, rays of light passing in the vicinity of a massive object are deflected from their initial direction by the amount of $\theta = 4GM/c^2b$, where $G$ is Newton's constant of gravitation, $M$ is the mass of the object, $c$ is the speed of light and $b$ is the ray's impact parameter. Rays with the same impact parameter converge at a focus where the light intensity is amplified by a factor of $2GM/(c^2\lambda)$, where $\lambda$ is the observing wavelength (Turyshev & Toth, 2017).

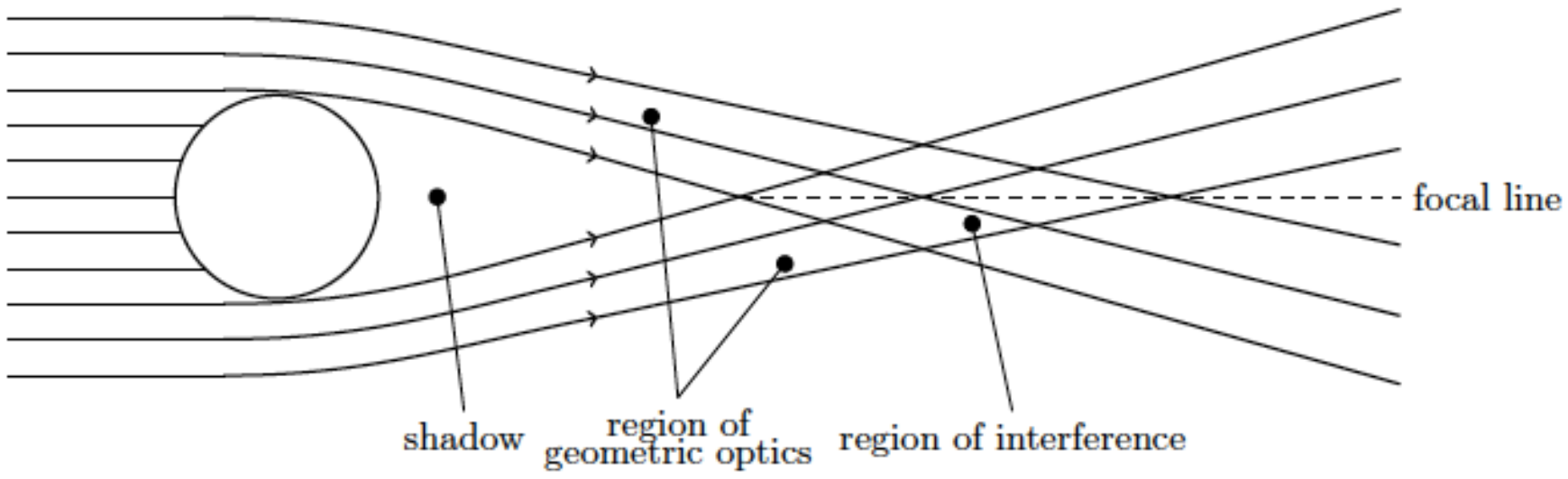

Figure 2 Solar gravitational lensing: Three different regions of space associated with a monopole gravitational lens: the shadow, the region of geometric optics, and the region of interference (from (Turyshev & Toth, 2017)).

Unlike an optical lens, a gravitational lens has spherical aberration, with the bending angle inversely proportional to the impact parameter of a light ray with respect to the lens. Therefore, such a lens has no single focal point but a focal line (Figure 2). Although all the bodies in the solar system may act as gravitational lenses (Turyshev, 2008), only the Sun is massive and compact enough for the focus of its gravitational deflection to be within the range of a realistic deep space mission. Its focal line begins at ~547.8 astronomical units (AU). A probe positioned beyond this distance from the Sun could use the SGL to magnify light from distant objects on the opposite side of the Sun (Eshleman, 1979).

The reason for the large amplification of the SGL is that, as a typical lens, the SGL forms a folded caustic[1] (Gaudi & Petters, 2001; Gaudi & Petters, 2002) in its focal area (see Figure 3.) As the wavelength of light is much smaller than the Schwarzschild radius[2] of the Sun that characterizes the gravitational lens, the wavefront in the focal region of the SGL is dominated by the caustic and singularities typical for geometrical optics. In reality, the geometric singularities are softened and decorated on fine scales by wave effects (Berry & Upstill, 1982; Berry, 1992). Although it yields divergent results, the geometric optics approximation may be used to predict the focal line (FL), and make qualitative arguments about the magnification and the size of the image. However, to design a telescope, one must address practical questions concerning the magnification, resolution, field of view (FOV), and the plate scale of the imaging system. These parameters are usually estimated by a wave optics approach and are needed to assess the imaging potential of the SGL.

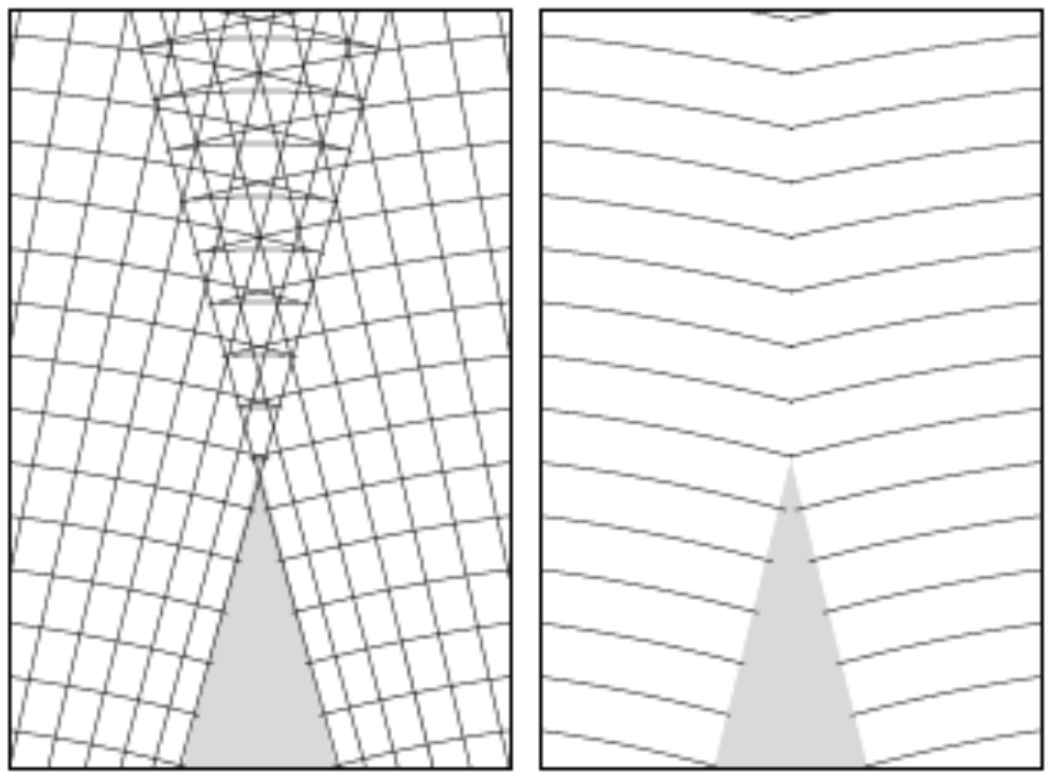

Figure 3 Folded caustic formed by the SGL (not to scale). Left: rays (thin straight lines) enveloping a cusped caustic and wavefronts, i.e., contours of travel time. Right: travel time contours as on the left, but showing only for first arrival at a particular point (from (Turyshev & Toth, 2017)).

Recently, we developed a wave-theoretical description of the SGL (Turyshev, 2017; Turyshev & Toth, 2017). We considered the propagation of electromagnetic (EM) waves in the gravitational field of the Sun, which is represented by the Schwarzschild monopole taken within the first post-Newtonian approximation of the general theory of relativity. We developed a wave-theoretical treatment for light diffraction in the field of a static gravitational mass monopole and considered the case of a monochromatic EM wave coming from a point source at a large distance from the monopole.

We obtained a solution for the EM field everywhere around the lens and especially in the immediate vicinity of its focal line, where geometric optics leads to diverging results. As anticipated, because of wave effects in the focal region, our wave-optical treatment is immune to singularities, allowing us to describe the optics of the SGL and understand its image formation properties. In contrast to models based purely on geometric optics, the new approach allows us to consider practical questions related to the design of a solar gravitational telescope, in part by permitting the use of traditional tools of telescope design.

[1]In optics, caustic is the envelope of light rays reflected or refracted by a curved surface or object, or the projection of that envelope of rays on another surface. Folds are one of the simplest caustics predicted by gravitational optics theory. When a source moves close to a fold caustic, two images will tend to merge with an arc-like geometry. The two merging images will vanish just after the crossing of the fold.

[2]The Schwarzschild radius (or the gravitational radius), $r_S = 2GM/c^2$, is the radius of a sphere such that, if all the mass of an object were to be compressed within that sphere, the escape velocity from the surface of the sphere would equal the speed of light. The Schwarzschild radius was named after the German astronomer Karl Schwarzschild, who calculated this exact solution for the theory of general relativity in 1916.

- Important features of the SGL (for $\lambda$ = 1 $\mu$m):
  - Major brightness magnification: a factor of $10^{11}$ (on the optical axis);
  - High angular resolution: ~0.5 nano-arcsec. A 1-m telescope at the SGL collects light from a ~(10km × 10km) spot on the surface of the planet, bringing this light to one 1-m size pixel in the image plane of the SGL;
  - Extremely narrow "pencil" beam: entire image of an exo-Earth (~13,000 km) at 100 l.y. is included within a cylinder with a diameter of ~1.3 km.
- Collecting area of a 1-m telescope at the SGL's focus:
  - Telescope with diameter $d_0$ collects light with impact parameters $\delta b \simeq d_0$;
  - For a 1-m telescope at 750AU, the total collecting area is: $4.37\times10^9$ m$^2$, which is equivalent to a telescope with a diameter of ~80 km…

Figure 4. Summary of the important properties of the SGL.

We showed that the light amplification of the SGL is $2GM/(c^2\lambda) \sim 10^{11}$ (for $\lambda$ = 1 μm) and its angular resolution is $\lambda/D \leq 10^{-10}$ arcseconds (0.1 nanorcsecond or nas), where $D$ is the solar diameter. The effective light collecting area, with a 1-m telescope placed at a heliocentric distance of ~750 AU, is equivalent to a diffraction-limited telescope with an aperture of ~80 km (
Figure 4.) Thus, the SGL may be used for direct megapixel imaging of faint objects.

Our results allowed us to study the point-spread function (PSF), resolution and FOV, as well as the evolution of these quantities at various heliocentric distances along the focal line. These quantities help us understand the unique properties of the SGL for imaging and spectroscopic investigations (Figure 5). The new knowledge of the optical properties of the SGL and the understanding of the current state-of-the-art in spacecraft technologies allowed us to propose for consideration a realistic mission to the focal area of the SGL with the objective of high-resolution imaging and spectroscopy of an Earth-like exoplanet.

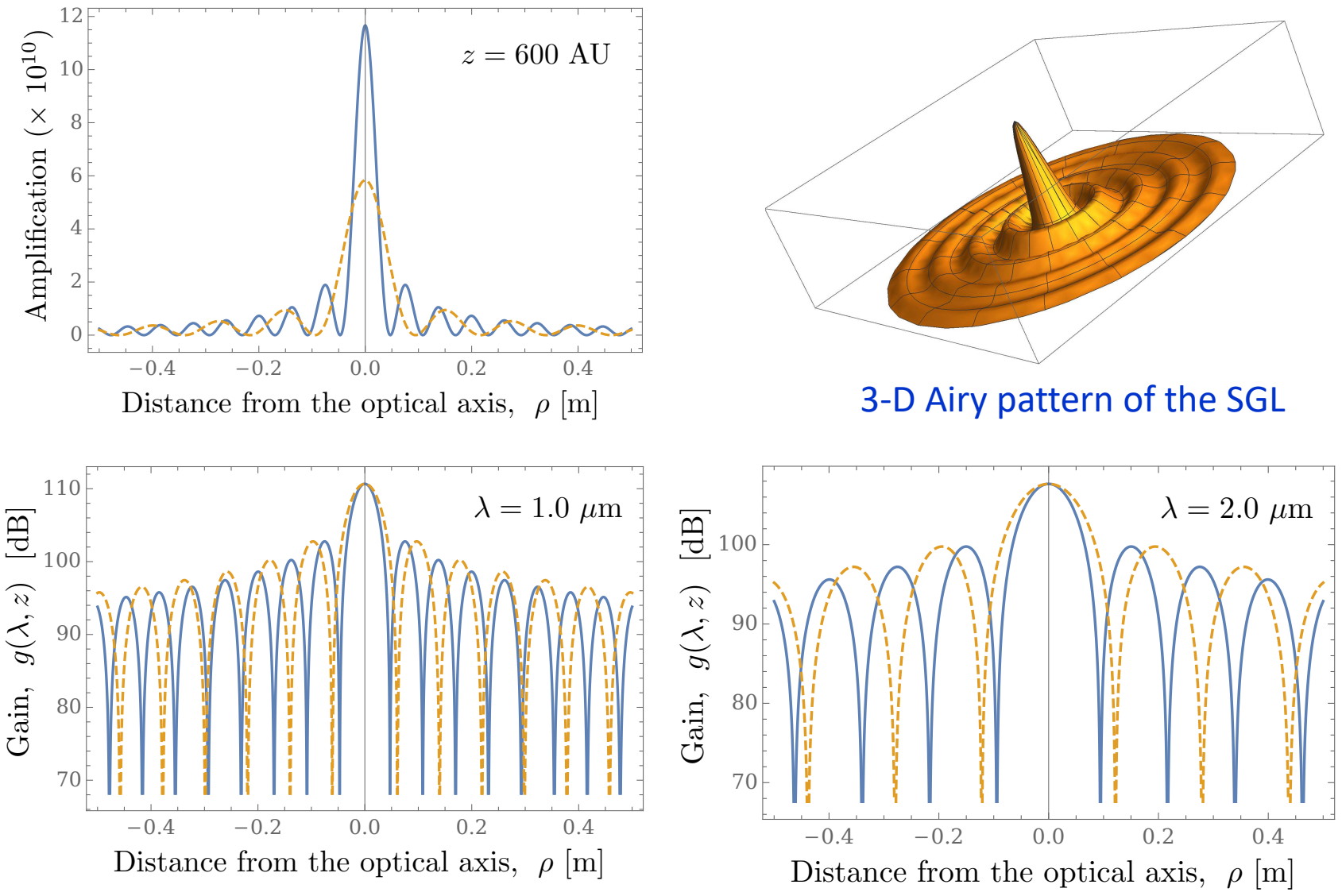

Figure 5. Optical properties of the SGL, from (Turyshev & Toth, 2017). Up-Left: Amplification of the SGL. Up-Right: Point spread function. Bottom: Gain of the SGL as seen in the image plane as a function of possible observational wavelength.

The exoplanet imaging mission concepts currently studied by NASA (Exo-C 2015, Exo-S 2015), capture the light of an unresolved Earth-like exoplanet as a single pixel. The limiting factor is contamination from the parent star ~0.1" from the planet. Clearly, a 3m telescope cannot detect an Earth at 10 pc (i.e., ~33 light years), much less at 30 pc. Even WFIRST with its telescope of 2.4-m is looking is only for exo-Jupiters at 10 pc. There is no concept for direct multi-pixel imaging of an exoplanet.

An Earth-like planet at 30 pc has an angular diameter of $1.4\times10^{-11}$ rad. A telescope with an aperture of ~80 km, which has the same light collecting area as the SGL, would barely resolve the disk of the planet. Resolving the planet with a thousand pixels across its diameter would require a telescope aperture of ~102,000 km (~$12R_\oplus$), which is obviously impractical. Building an imaging interferometer with a set of such baselines is similarly not feasible (Figure 6).

- Overcoming the issue of a small target size:
  - Consider an exo-Earth @ 30pc (100 l.y.) is ~$1.4\times10^{-11}$ rad;
  - A diffraction-limited telescope needed to resolve an object with this size at such distance must have a diameter of ~76 km;
    - But, even this telescope would barely resolve the disk of the planet.
  - To resolve the planet with 1,000 pixels one needs a telescope with a diameter of $7.6\times10^4$ km (or ~12 $R_\oplus$), which is impractical...
    - An imaging interferometer with a set of such baselines - not feasible.
  - Even more challenging is the integration time needed to reach SNR=10:
    - a 50m telescope would need an integration time of $t\sim10^6$ years (zodi);
    - with SGL's light amplification (~$2\times10^{11}$) we could do the job in ~7 weeks.
- Solving the parent start light contamination issue:
  - Current exoplanet-imaging concepts detect light of a planet as a single pixel. Contamination from the parent star (~0.1" off the planet) is a major problem;
    - Due to the high angular resolution of the SGL (~0.5 nas), the parent star is resolved from the planet with its light amplified 0.01 AU away from the optical axis, making the parent star contamination issue negligible.

Figure 6. The SGL enables direct multi-pixel imaging: comparison with the conventional techniques – a telescope and an interferometer.

The discovery of numerous exoplanets by the Kepler telescope, including those that may be Earth-like (Torres et al, 2015), created interest in methods to image these distant worlds. The success of the Voyager-1 spacecraft, operating at a distance of nearly 140 AU from the Sun, demonstrates the feasibility of long duration deep space missions beyond the outer solar system, including regions where images are formed by the SGL. The idea of using the SGL for direct megapixel high-resolution imaging of an object of extreme interest, such as a habitable exoplanet, was extensively discussed in a recent study at the Keck Institute for Space Studies (KISS) (see (Stone et al., 2014)).

Progress in technologies for long duration spaceflight is required to reach large heliocentric distances. Various aspects of those were demonstrated by the success of the Voyager 1 and 2, Pioneer 10 and 11, and New Horizons missions. With the recent progress in the development of highly capable small spacecraft, electric propulsion techniques, optical transponders, optical quality foldable mirrors, etc., the search for exoplanets now stands ready to benefit from these spacecraft-related technologies.

We may now consider missions to the very deep regions in the outer solar system (Alkalai et al., 2017). Clearly, such missions must capture the public's imagination and support during the anticipated destination transit times. Finding alien life on a distant exoplanet is an exciting objective for such a mission. Recognition of such a unique opportunity led to our NIAC 2017 proposal (Turyshev et al., 2017) to study a mission to the focal region of the SGL with an objective of direct high-resolution imaging and spectroscopy of a habitable exoplanet.

This is the Final Report for our NIAC 2017 Phase I investigation, structured as follows:

In Section 2 we discuss our current understanding of the mission and instrument requirements. Our imaging approach is based on the knowledge of the optical properties of the SGL. We consider the concept of operations, instrumental design, as well as the rotational and direct deconvolution approaches. We also present our considerations on the target selection and anticipated properties.

Section 3 highlights our mission design studies conducted to explore the feasibility of a mission capable of operating at the focal region of the SGL. We discuss the study approach and relevant mission tradeoffs. We consider critical elements of the SGL mission design and consider heliocentric ranges for the mission. We present and discuss various mission concepts that were considered during the study, including a single spacecraft, a "string-of-pearls" approach, and a mission concept based on solar sail technology.

In Section 4 we discuss the results that we obtained and present our recommendations.

# 2 IMAGING WITH THE SOLAR GRAVITATIONAL LENS

In the past, only the amplification properties of the SGL under a set of idealized physical conditions were investigated, considering only the gain of a combined receiver consisting of a large parabolic radio antenna, at the focus of which there was a single pixel detector situated on the focal line of the SGL (Eshleman, 1979; Maccone, 2002, 2009, 2010, 2011). The SGL's imaging properties, where the image occupies many pixels in the immediate vicinity of the focal line, are not yet fully explored (except perhaps for some introductory considerations of geometric raytracing (Koechlin et al, 2005; Koechlin et al, 2006)), especially in the context of a deep-space mission. In addition, the SGL's potential for high-resolution spectroscopy should also be studied. Our NIAC Phase I proposal is the first dedicated effort in this direction.

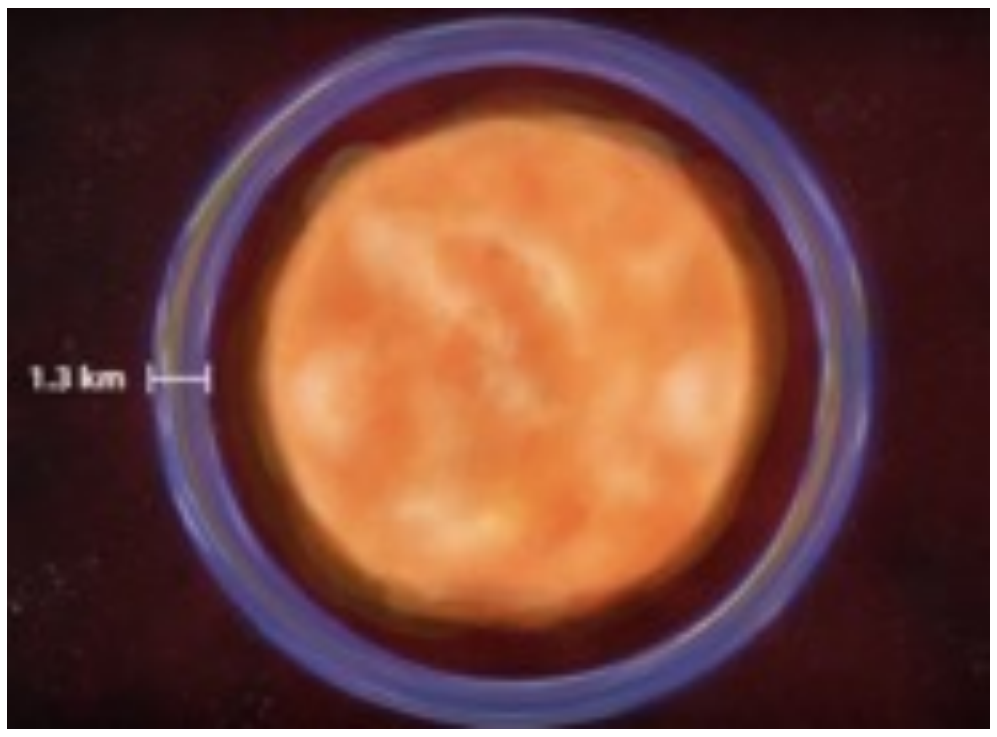

Figure 7. Einstein ring with thickness of 1.3km around the Sun formed by an exo-Earth (not to scale!). From (DeLuca 2017).

As light from distant sources passes around the Sun, deflection by the solar gravitational field bends photon trajectories (Figure 2), resulting in a magnification of the light's intensity. Spacecraft moving along the focal line (FL) could take observations, communicate, and send data back using equipment typically used for interplanetary missions (Cesarone et al., 2014). This is why the principal objective of our NIAC study was to examine the feasibility of such a project and to document the expected outcomes of such a deep space mission.

## *2.1 Concept of operations*

As seen from a telescope at the FL of the SGL, light arriving from an exoplanet at the imaging telescope forms an annulus called the Einstein ring. Given the plate scale, for an exoplanet with the diameter of $D_{\rm ep}$ situated at a distance of $z_{\rm ep}$, from the Sun, its entire image is compressed by the SGL into a region with a diameter of $\sim D_{\rm ep}\,(z/z_{\rm ep})$ in the vicinity of the focal line, where z is heliocentric observing distance. For an Earth-like exoplanet at $z_{\rm ep}$ of 100 ly away and at an observing distance, $z$, beyond 650 AU this will be a cylinder with a diameter of ~1.3 km (z/650AU) (see Figure 7). The thickness of the ring of 1.3 km is not resolved by a 1-m observing telescope at the focal region of the SGL. All the imaging and spectroscopic information is in the variable ring's brightness.

The diameter of the exo-Earth image which is compressed by the SGL is equal the thickness of the Einstein ring that occupies a region around the edge of the Sun (Figure 8). As the angle of the gravitational deflection is inversely proportional to the solar impact parameter, the radius of this ring $b$ slowly increases with heliocentric distance $z$, as $b \sim R_\odot\sqrt{z/z_0}$, where $R_\odot$ is the solar radius and $z_0$ = 548.7 AU is the closest distance to the focal area of the SGL. Thus, the further the helio-centric distance, the less will be the contribution form the corona. Although the Einstein ring is much dimmer than the Sun, using a modest coronagraph (i.e., one with ~$10^6$ suppression) to block the brightness of solar disk down to that of the solar corona, makes it possible for light from the exoplanet to be detected by the telescope.

We define the radius of the Einstein ring formed by the SGL at a given heliocentric range as a ring at a certain the distance from the observed center of the Sun to the center of the ring, and the width of the ring as the distance from the inner to the outer edges of the ring (see in Figure 8.)

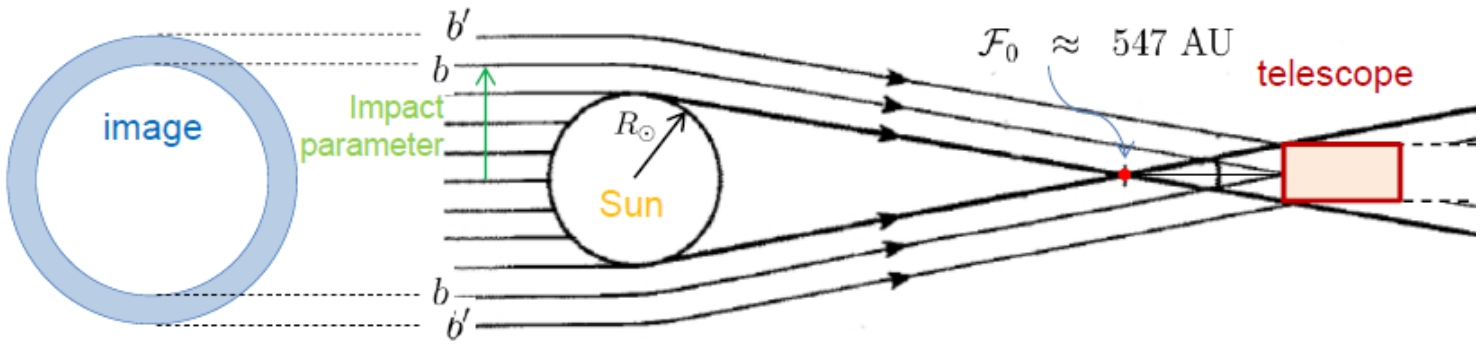

Figure 8. Imaging of an exo-Earth with solar gravitational lens. The exo-Earth occupies (1km×1km) area at the image plane. Using a 1m telescope as a single-pixel detector provides a 1000 × 1000 pixel image.

Consider the basic geometry of the problem: At ~550 AU the Sun subtends an angle of ~3.5". For the wavelength of $\lambda$ = 1 μm, the diffraction-limited size of a 1-m telescope has a beam size of $\lambda$/(1 m) ~ 0.1". Most light within the Einstein ring comes from ~10 km wide region on the exoplanet's surface. However, because of the large spherical aberration of a gravitational lens (Turyshev & Toth, 2017), light from adjacent areas on the exoplanet also contributes to the annulus, albeit to a lesser extent. The knowledge of the optical properties of the SGL (e.g., PSF, etc.) will be important for deconvolving the image from the distortions introduced by these spherical aberrations.

We consider an exoplanet that is similar to our Earth, with a diameter of ~13,000 km. We assume that it orbits a star similar to our Sun, situated at 100 light years (~30 pc) from us. For a spacecraft on the optical axis of the SGL, the image of such an exoplanet is compressed (plate scale) into a cylinder with a diameter of ~1.3 km ($z/z_0$), centered on the optical axis. To develop a multi-pixel image of an exo-Earth with a high resolution of ($10^3 \times 10^3$) pixels, the spacecraft would have to scan this 1.3 km × 1.3 km area one pixel at a time. Alternatively, one could envision a constellation of several apertures used for this purpose.

In the context of the optical properties of the SGL, the effect of the solar corona was investigated (Turyshev & Andersson, 2003) using a geometric optics approach. It was shown that propagation of radio waves in the immediate vicinity of the Sun is significantly affected by the solar plasma, which effectively pushes the focal area of the SGL to large heliocentric distances. This is happen-ing because the electron plasma of the solar corona has a negative index of refraction. The phase of a light ray is delayed by the amount that is proportional to the square of the ratio of the wave-length to the impact parameter, ~$(\lambda/b)^2$. For long RF wavelengths (i.e., S-band) and short impact parameters, the solar corona introduces a very large delay, so that the EM signal is effectively stopped by the solar corona. Also, the entire wavefront of an EM wave appears to be deflected outwards. Gravity, however, bends the wavefront inwards. So, the plasma and gravity counter each other. Again, for RF signals and short impact parameters, the effective deflection angle is reduced, pushing the focal area to higher heliocentric ranges (Turyshev & Andersson, 2003). In contrast, the propagation of optical waves is not significantly affected by the plasma.

Nevertheless, effects of the radial/azimuthal plasma density of the solar corona (Turyshev & Andersson, 2003) on the structure of the lensing caustic must be taken into account, including analysis of second order effects and the chromatic structure of the caustic. Some of these investigations are already completed. We studied (Turyshev, 2018a) the combined influence on the optical properties of the SGL due to the static monopole gravitational field of the Sun, modeled within the first post-Newtonian approximation of the general theory of relativity, and the solar corona, as a function of the heliocentric distance and modeled using a generic power law model.

To prepare for quantitative analysis of the mission design, we studied the physical properties of the SGL. This gave us the understanding of the image formation process in the context of a space mission operating beyond 550 AU. We addressed the impact on the optical properties of the SGL from the solar corona (Turyshev, 2018a). We also studied the contribution of the Sun's extended structure, where its monopole gravity field is augmented by the field produced by the solar gravitational quadrupole and solar rotation (Turyshev, 2018b).

In this regard, we considered the propagation of monochromatic EM waves near the Sun and developed a Mie theory needed to account for refractive properties of the free-electron plasma in the immediate solar vicinity. This resulted in an enhanced theory of the SGL, accounting for the contribution of the static part of the solar plasma to the optical properties of the SGL. Because of their smaller magnitude, temporal variations are less significant in the PSF and also are expected to be averaged out during integration.

We demonstrated that although the solar corona's contribution may be important in radio frequencies, it is negligible at optical frequencies. Furthermore, we demonstrated that solar oblateness and the static component of the solar plasma result in a nearly constant phase shift for each of the individual light rays that form the image (Turyshev 2018a, 2018b). Therefore, to first order, these contributions does not affect the magnification of the SGL. To remove these effects completely, we can use standard deconvolution tools and the models developed for those purposes.

An outcome of these recent papers was the recognition that one may neglect the effect of the solar corona on the optical properties of the SGL, as its PSF is minimally affected by the corona. In fact, solar corona has insignificant effect on the trajectory of a light ray, however, it's brightness is important for imaging as it affects the SNR. These results were important in our efforts to study the instrumentation design for a mission to the SGL as they allowed us to concentrate on the brightness of the solar corona and its mitigation with an on-board coronagraph.

We also considered the impact of solar motion with respect to the solar system barycentric coordinate reference system (BCRS, see (Turyshev & Toth, 2013)). Figure 9 shows the position of the center of the Sun as it moves with respect to the BCRS under the gravitational influence of the other planets, primarily Jupiter and Saturn (Figure 10). The Sun's position slowly varies with a period of ~12 years. Although this motion is slow, it changes the direction of the focal line. Any mission that would operate at the SGL must have enough propulsion capability to follow the articulation of the optical axis of the SGL induced by the motion of the Sun. In this respect, it makes sense to introduce a concept of the instantaneous focal line of the SGL – which is the line connecting the center of the exoplanet and the center of the Sun as it moves with respect to the solar system's BCRS. This concept is helpful to study the imaging concept with the SGL.

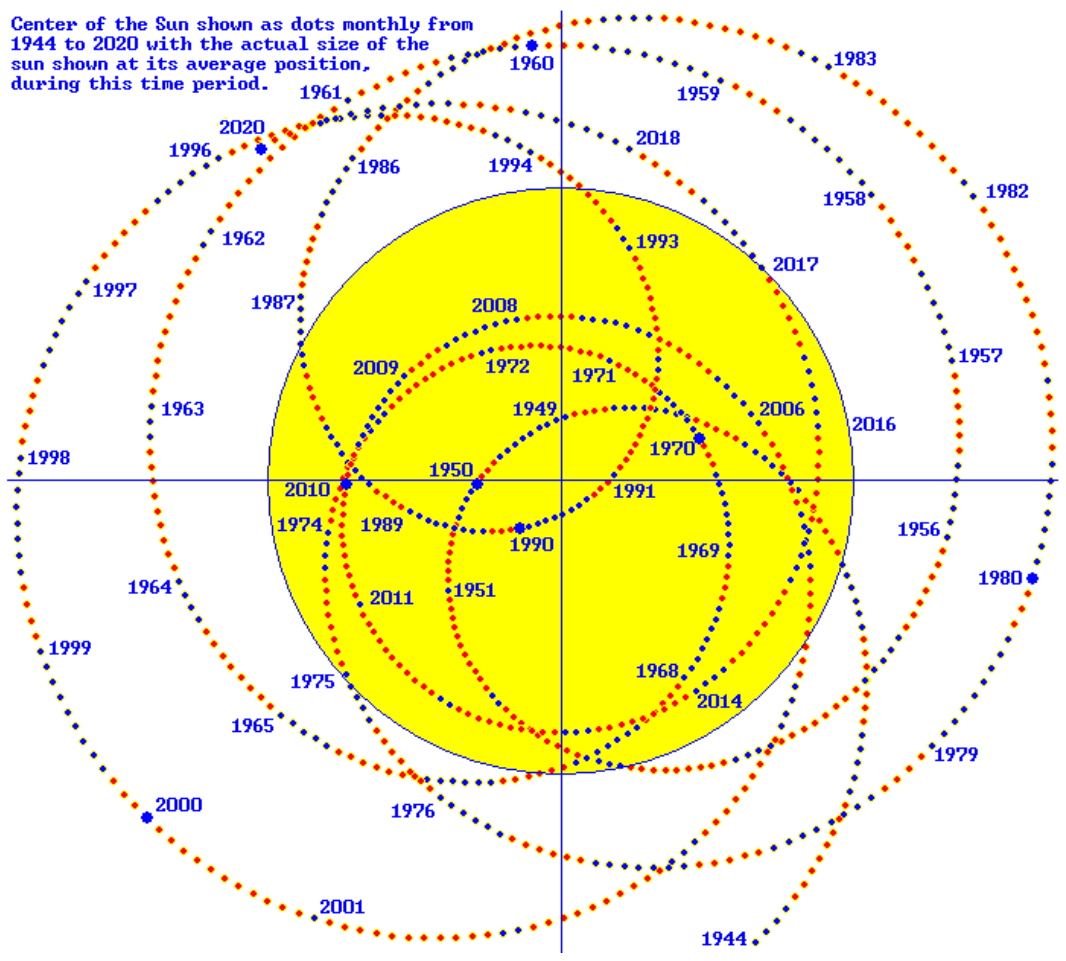

Figure 9. The solar wobble: Center of the Sun shown as dots monthly from 1944 to 2020 with actual size of the Sun shown at its average position, during this time period.

- Pointing precision (between three objects):
  - Needs to be maintained to ~ few $\mu$as for proper operation of the SGL.
  - Knowledge is needed at 1 $\mu$as level, control is at the ~100 $\mu$as.
  - The motion is unfortunately complex (1m of motion at 600AU ~ 1 $\mu$as of angle seen from Earth)
- Simple motions (straight lines):
  - Motion of the target star around the galaxy; the Sun around the galaxy
- More complex motions:
  - Motion of the exoplanet around its host star (Keplerian)
  - Motion of our Sun around the solar system barycenter.
    - Dominated by the orbits of Jupiter, Saturn.
    - Jupiter => 75 million m motion of the Sun (12yr orbit)
    - Saturn => 50 million m of motion of the Sun. (29 Yr)
    - Earth => 450,000 m 1yr
- Propulsion system must compensate for the reflex motion of the Sun
  - Due to most of the planets in the solar system. (perhaps many of the big asteroids in the main belt) Uranus and Neptune's motion over a short time may be just a straight line (need to calculate for sure).

Figure 10. Various motions affecting an SGL mission. Spacecraft would have be able to either measure or control these motions.

Accounting for the motion of the Sun in the solar system BCRF (Figure 9), consider the plate scale. The SGL reduces the image of the exo-Earth by a factor of ~6,000. An orbital radius of 1 AU becomes ~$2.25\times10^4$ km and an orbital velocity of 30 km/s translates into 5 m/s. Solar gravity accelerates the Earth at 6 mm/s$^2$. Consequently, the imager spacecraft needs to accelerate at ~1 μm/s$^2$ to move in a curved line mimicking the motion of the exoplanet. Even if the probe mass is ~1,000 kg, the corresponding force is 1 mN, which may be achieved with electric propulsion. It is even more reasonable for a ~100 kg notional spacecraft, assumed for this study. Whether it is feasible for interplanetary propulsion, electric propulsion may be sufficient for maneuvering to sample the pixels in the Einstein ring needed for imaging.

We investigated the attitude control of the SGL spacecraft. For that we considered the use of 3-axis stabilized spacecraft with a few microarsecond pointing knowledge and stability in combination with a set of laser beacons in the inner solar system. We studied the motion in the image plane needed to sample ($10^3\times10^3$) image pixels. This analysis allowed us to formulate and study the image reconstruction requirements that led us to formulate the key mission and instrument requirements. The also helped us to analyze image formation processes and to derive realistic mission requirements with relevant architecture trades to be discussed in Section 3.

## *2.2 Instrumentation design*

The photometric gain of the SGL, its high angular resolution and strong spectroscopic SNR of $10^4$ in 1 sec (Shao et al., 2017), suggest that the primary instrument should implement a miniature diffraction-limited high-resolution spectrograph, taking full advantage of the amplification and differential motions (e.g. exo-Earth rotation).

- The instrument:
  - A diff.-limited high-resolution spectrograph, enabling Doppler imaging techniques;
  - Within 30 pc, a 1 AU HZ will project to ~0.033", so such a planet maybe detectable with a high contrast ($10^{-10}$) at 3λ/d coronagraph with a 9m telescope;
  - Given the rapid development of coronagraphic capabilities, we can assume that direct imaging will provide spectro-photometric characterization of the exo-Earth;
- The SGLF telescope need a coronagraph to block the light from our Sun:
  - A conventional coronagraph would block just the solar light, but we want the coronagraph to transmit light only at the Einstein ring (where the planet's light is).
  - At 1 μm, the gain of the SGL is ~110dB (27.5 mag), so an exoplanet, which is 32.4 mag object, will become a ~4.9 mag object;
  - When averaged over a 1m telescope (the gain is ~$2\times10^9$), it would be 9.2 mag, which is sufficiently bright (even on the solar background);
  - To derive an image with the SGLF, including solar corona brightness (the parent star will be resolved), zodiacal light, instrument, and s/c systematics;
- Perhaps several small spacecraft?
  - We could rely on a swarm of small spacecraft, lunched together each moving at a slightly different trajectory parallel to the optical axis.

Figure 11. The instrumental design for imaging with the SGL.

Consider the basic geometry: We estimate that within 100 light years, a 1 AU habitable zone will project ~0.033" (with dependence on spectral type). Such a habitable zone may be detectable with an ultra-high contrast (i.e., ~$10^{-10}$) at 3λ/d coronagraph with an 18-m telescope in space. (Finding the exoplanet directly while relying on traditional optical techniques will be very hard.) However, once the target is known, then, due to the progress in coronagraphic capabilities, we can assume that direct imaging with the SGL will provide spectrographic and photometric characterization of the exoplanet (Figure 11).

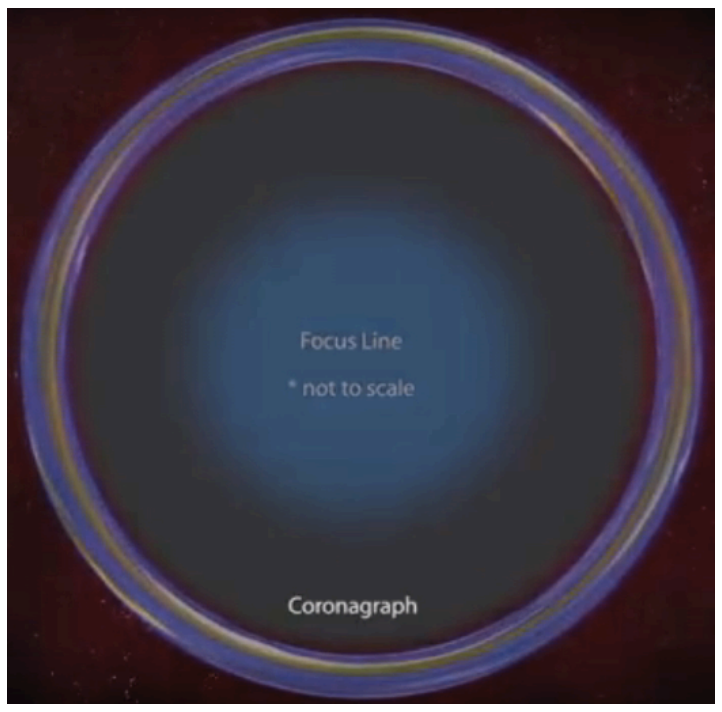

Figure 12. Using a solar coronagraph: Einstein ring visible on the background of the solar corona. From (DeLuca 2017).

As the instrument ultimately determines the size of the spacecraft and its motion in the image plane, we began our efforts with the instrument design. Specifically, we addressed the coronagraph design as a part of our study. We initially assumed a coronagraph in the shape of the annulus that corresponds to the Einstein ring around the Sun (as seen from the focal area at a given heliocentric location). However, this turned out to be unnecessary. A better approach is to require the coronagraph to block the solar disk to the level set by the solar corona brightness at a given position of the Einstein ring (Figure 12). Although the thickness of the ring stays nearly the same, the separation of its inner edge from the edge of the Sun increases as heliocentric distances increase. Reducing solar brightness below of the solar corona at a particular heliocentric distance is not necessary.

The instrument must be able to sample various spots on the image plane as it is moving along the optical axis at a high heliocentric speed. These spots are located up to ~650 m (the radius of the cylinder occupied by the exo-Earth image) from the optical axis.

As a result, we require the SGL telescope to have a coronagraph capable of blocking the solar disk to the level set by the solar corona brightness at a particular separation of the Einstein ring from the Sun. The solar corona brightness as a function of the solar separation is well known and is shown in Figure 13.

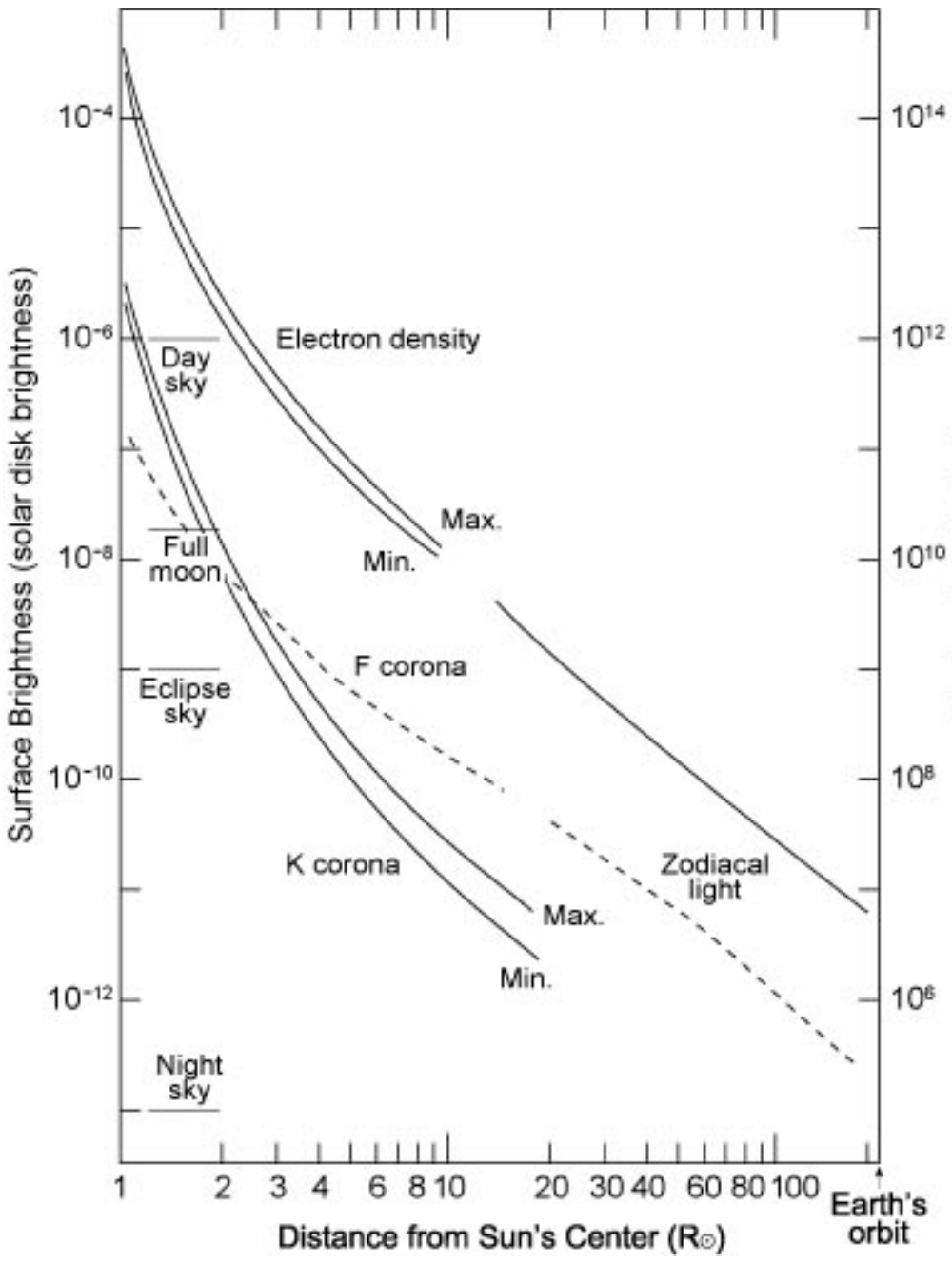

Figure 13. Solar corona brightness (Lang, 2010).

At 1 μm, the light amplification of the SGL is ~2 × $10^{11}$ (which is equivalent to additional 27.5 stellar magnitudes), so an exoplanet, which is initially seen as an object of 32.4 mag, becomes a ~4.9 mag object. When averaged over a 1-m telescope, the light amplification is reduced to ~2×$10^9$ (Turyshev & Toth, 2017)), resulting in a brightness of 9.2 mag, which is sufficiently bright (even on the solar corona background). To capture an image with a telescope positioned at the SGL's focal area, in addition to the light coming from the exoplanet, the telescope will collect light from solar corona, residual solar light (not filtered by the coronagraph), and the zodiacal light. Light contributions from the parent star were not considered, as its light is focused thousands of kilometers away from the optical axis. Therefore, for this study the photometric flux from the parent star is immaterial.

To validate our design assumptions, we performed a preliminary coronagraph design and related simulations. The need to suppress the Sun's light by a factor of $10^{-6}$ when imaging with the SGL is significantly less demanding than modern-day exoplanet coronagraphs, which aim to suppress the parent star's light by a factor of $10^{-10}$ to detect an exo-Earth at least as a single pixel. However, the design for the SGL's coronagraph is not trivial. The main difference between the SGL coronagraph and those used in modern planet-hunting approaches is that the parent star is not resolved, however, for the SGL concept the Sun is an extended object. The coronagraph bears greater similarities to classical coronagraphs used in solar astronomy rather than to the exoplanet coronagraphs being studied for exoplanet imaging. A traditional solar coronagraph uses sharp edge masks in both the focal plane and pupil plane. In our case, a more complex mask such as a classic Lyot coronagraph may be needed. Specifically, a mask flat within the radius of the Sun and of a Gaussian profile soft edge achieves better rejection than a mask with a sharp edge.

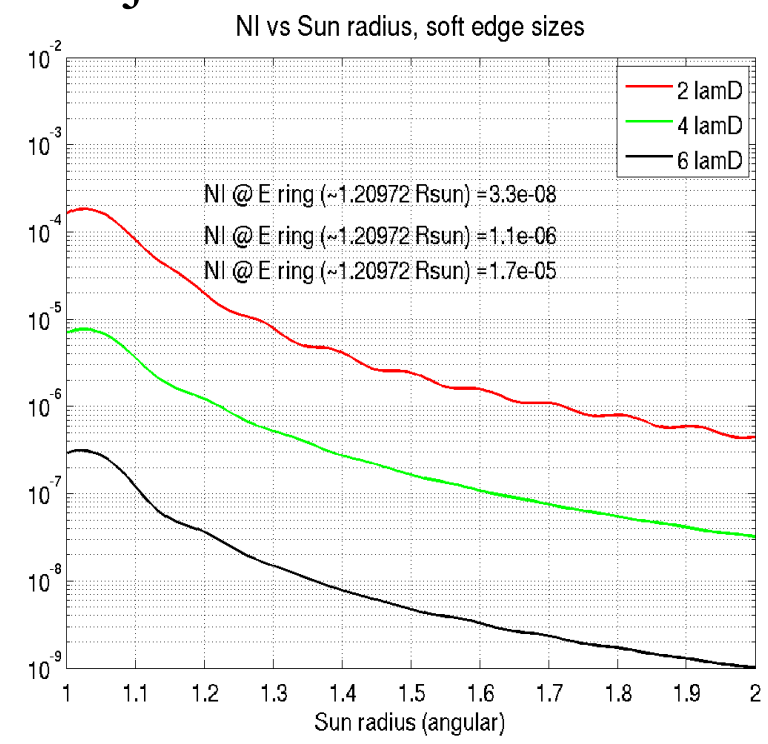

Figure 14. The FWHM of Gaussian soft-edge has a great impact on light suppression ability of the coronagraph.

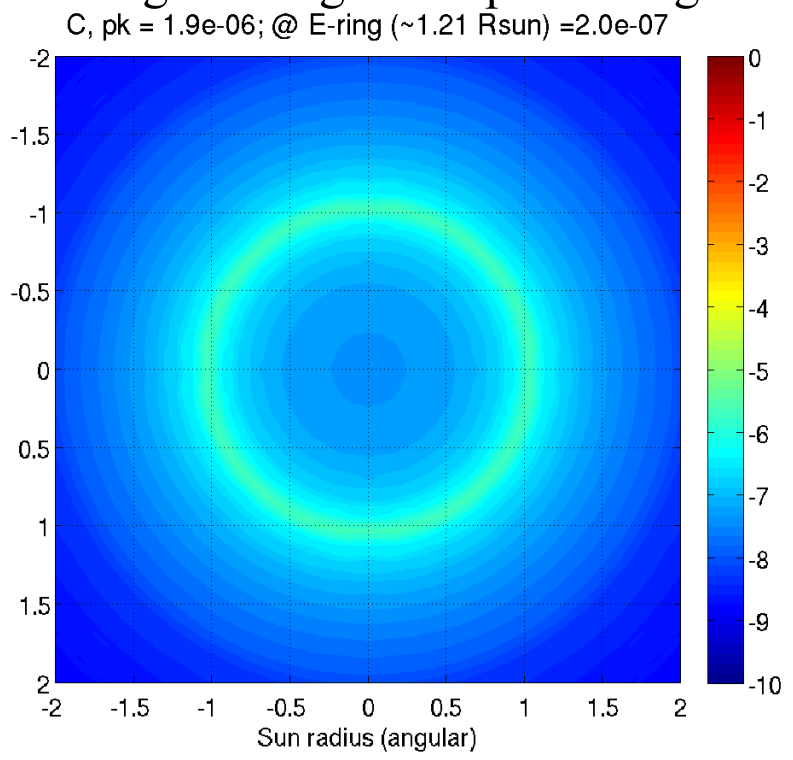

Figure 15. Simulated coronagraph performance showing the solar light suppression at the level of $2 \times 10^{-7}$, sufficient for imaging with the SGL.

The performance of the coronagraph was evaluated with a Fourier-based diffraction model. The Sun is modeled as a collection of incoherent point sources (of uniform brightness and monochromatic visible wavelength) with its corona (of $1/r^3$ power law radial profile). Design parameters include telescope size, SGL distance, occulter mask profile, and Lyot mask size. The FWHM of the Gaussan soft edge, for example, has a significant impact on the coronagraph's performance, as illustrated in Figure 14. Our initial results based on this simulation suggest that we have to be at the distances of 800–900 AU from the Sun and use a 2-m telescope in order to suppress the Sun's light to a level below the solar corona. The 2-m size of the telescope is driven by the limitations to the optical throughput of the system imposed by the coronagraph (most loss comes from the Lyot stop mask in the current design). This opens up a trade space to be explored during Phase II. Nevertheless, our baseline approach of having a 1-m telescope 600-850 AU range still feasible.

We were able to achieve a total planet throughput of ~10%, which may be improved. Figure 15 shows the contrast at the image plane after the coronagraph of the diffracted light from the Sun and the flux from the solar corona. Contrast is defined as brightness normalized to peak brightness without the coronagraph. At a contrast of $2 \times 10^{-7}$ at the E-ring, light leakage from Sun is roughly five times less than from the solar corona, sufficiently above the needed suppression.

These initial results gave us confidence that we can have a realistic instrument design, which satisfies the objectives for imaging with the SGL. We also identified a set of trade parameters that could lead to the optimal design. For example, additional resource for aperture size reduction would be to go to larger heliocentric distances beyond 1,000 AU. At those distances, the Einstein ring is further away from the Sun, which results in an increase of its relative brightness compared to that of the solar corona. This option opens up a design tradeoff: offloading some of the difficulties in achieving the required optical performance onto the mission architecture and design needed to reach greater heliocentric distances (Figure 16).

Alternatively, we may consider an external coronagraph (e.g., a starshade-like coronagraph similar to that considered for WFIRST) or a hybrid internal /external coronagraph. The use of a starshade could allow for significant reduction in aperture size of the SGL telescope, providing the opportunity for a swarm of small spacecraft to be considered for the task. Study of this tradeoff will be included in the Phase II proposal.

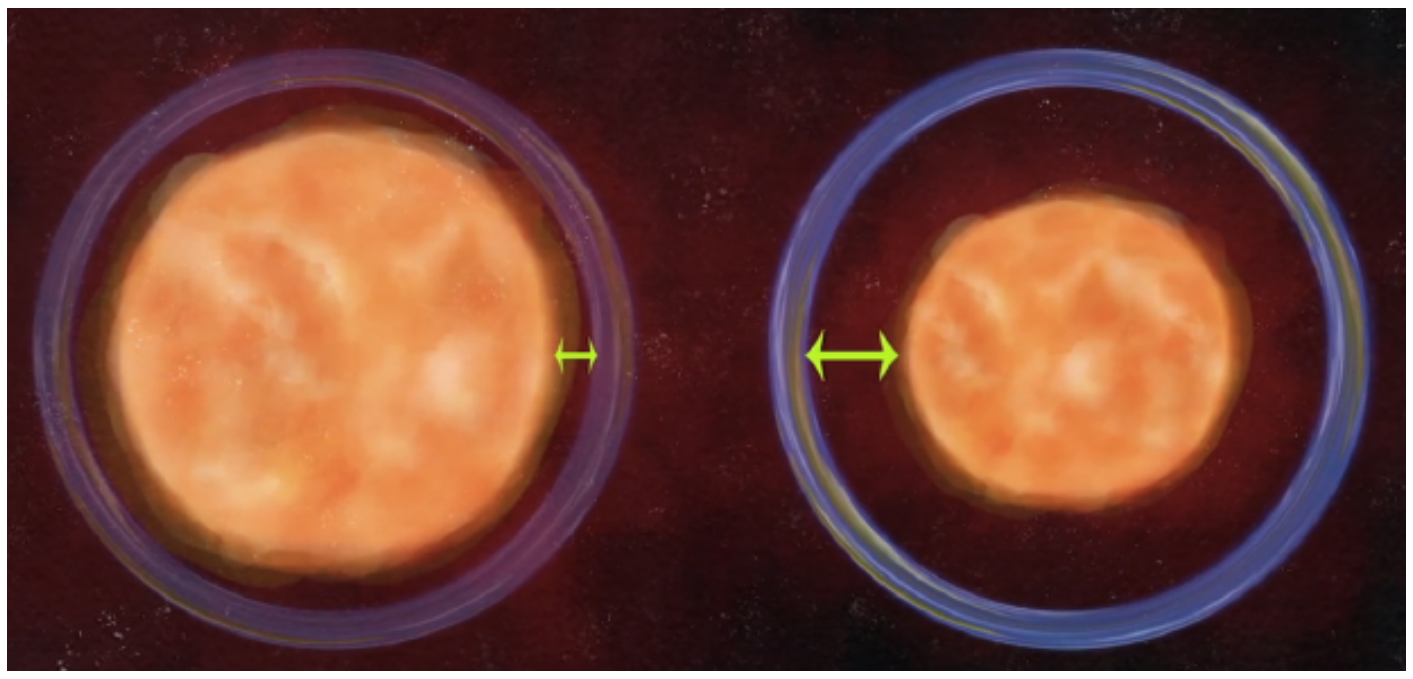

Figure 16. At larger heliocentric distances (from left to right), the Einstein ring is further separated from the Sun. Thus, less solar corona would influence the imaging process (DeLuca 2017).

We considered the tradeoffs between a traditional telescope vs. a microsatellite system. A small telescope would have limited capabilities, but it may open the possibility of sending multiple spacecraft. In fact, we could also devise an instrument that would rely on a swarm of small spacecraft, perhaps even launched together but each moving at a slightly different trajectory parallel to the instantaneous principal optical axis. Such an instrument would rely on the light collection capabilities enabled by a formation flying architecture. This would allow us to verify our initial assumption that the instrument should implement a miniature diffraction-limited high-resolution spectrograph, taking full advantage of the SGL amplification and differential motions. This tradeoff will also be explored during our Phase II study.

### *2.3 Imaging with the SGL*

The key mission and instrument requirements are driven by the process of image formation in the context of a realistic space mission capable of operating at large heliocentric distances. With the newly developed wave optics treatment for the SGL (Turyshev, 2017; Turyshev & Toth 2017; Turyshev, 2018a), we were able to describe the process of image formation by the SGL. We know the structure and properties of the caustic formed by the SGL and can study its spatial, temporal and spectral characteristics. This new knowledge was helpful in guiding our mission design.

Based on plate scale, the image of an exo-Earth at 30 ps is compressed by the SGL to a cylinder with a diameter of ~1.3 km in the immediate vicinity of the FL. (This dimeter corresponds to the Einstein ring around the Sun with a thickness of 1.3 km.) Imaging exo-Earth with $(10^3\times10^3)$ pixels requires moving spacecraft in the image plane in steps of $1.3\text{km}/10^3 \sim 1.3$ m while staying within the instantaneous cylinder with the diameter of ~1.3 km. So, each ~1 m pixel in the image plane correspond to $13\times10^3$ km/$10^3\sim$ 10 km pixel sizes on the surface of the planet.

The challenge comes from recognizing the fact that the PSF of the SGL is quite broad (see Figure 5), falling off much slower than the PSF of a typical lens. For any particular pixel on the image plane, this leads to admixing the light from many surface pixels adjacent to that on the instantaneous FL into that one image pixel. This admixing results in a significant image blurring.

To overcome this challenge, imaging must be done on a pixel-by-pixel basis by measuring the brightness of the Einstein ring at each of the image pixels. With the knowledge of the PSF one can use modern deconvolution tools that allow for efficient reconstruction of the original image. However , for this process to work, a significant signal to noise ratio (SNR) is required. Luckily, the SGL's magnification leads to SNR of over $10^3$ in 1 sec, sufficient for a nearly noise-less deconvolution. (We discuss our simulations of such an image reconstruction in Secs. 2.3.1 and 2.3.2.)

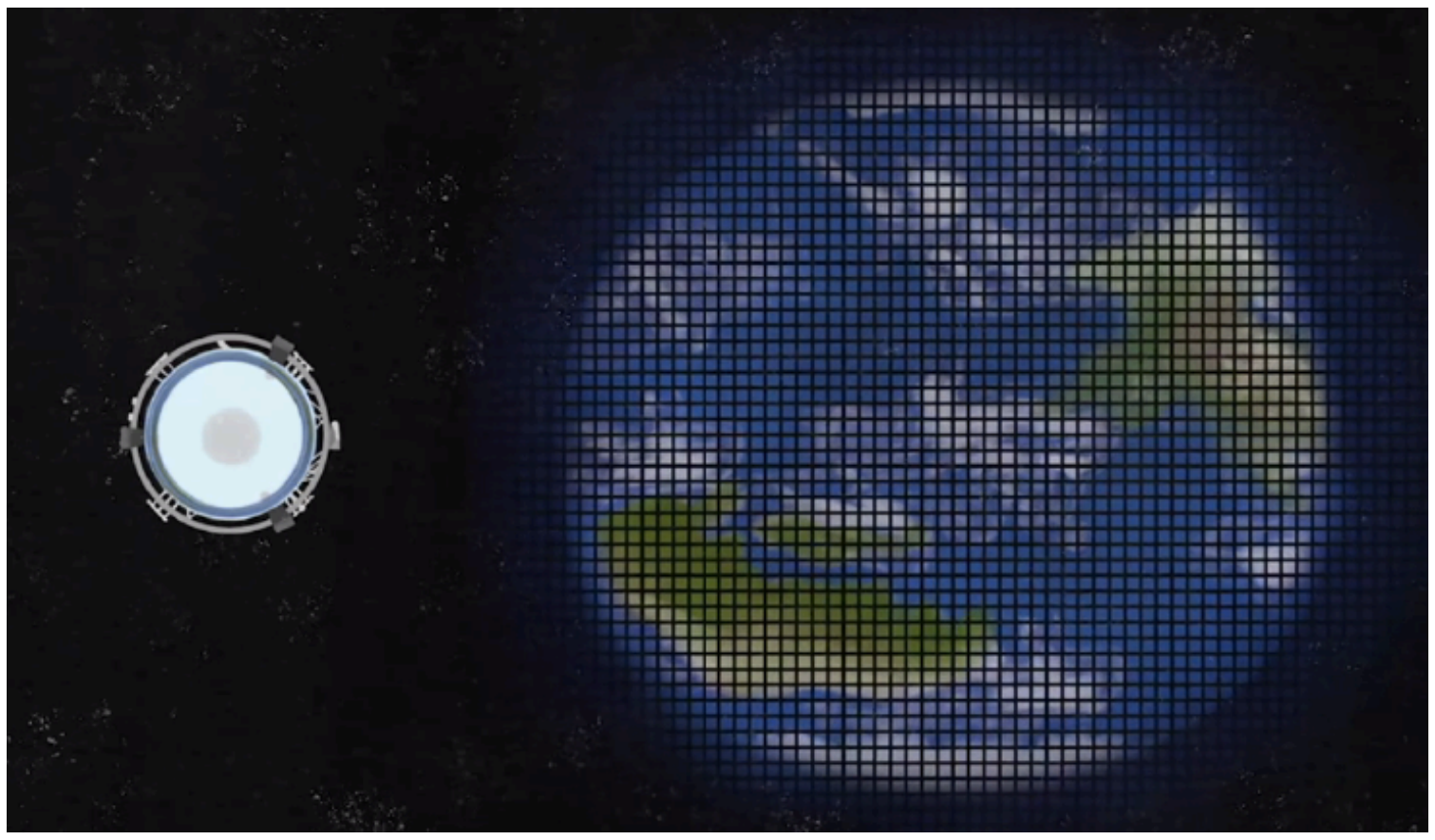

Figure 17. Imaging with an SGL telescope is on a pixel-by-pixel basis. At each pixel, we record the pixel's position and the photometric signal, and then use deconvolution tools to reconstruct the image. Picture from (DeLuca 2017).

Initial acquisition of an orbiting and rotating exoplanet may be done (as it moves across the FOV) using the recently developed technique of synthetic tracking (Shao et al. 2014, 2015), capable to finding fast-moving, dim targets.

- Imaging is done on a pixel-by-pixel basis:
  - The image of an exo-Earth occupies ~(1.3km×1.3km) area from the optical axis.
  - Each pointing corresponds to a different impact parameter: 1 image ⇔ 1 pixel.
  - Between the adjacent pixels the impact parameter changes, brings light from adjacent surface areas on the planet ⇒ a raster scan moving the spacecraft;
  - To build a ($10^3$×$10^3$) pixels image, we would need to sample the image pixel-by-pixel, while moving in the image plane with steps of ~1 km/$10^3$ = 1 m:
    - Pointing: Inertial navigation and 3 laser beacon spacecraft in heliocentric orbit in the plane of the Einstein's ring (for precision pointing & comm).
  - Contamination from the parent star is negligible for an SGL scenario.
- Exoplanet imaging requires several key technologies that are challenging:
  - determination of an exoplanet astrometric orbit at ~10 nas,
  - motion & stabilization of the s/c over millions of pointings with limited power.
- Perhaps even spectroscopy or even spectro-polarimetry of the exoplanet?
  - Potentially a spectrally resolved image over a broad range of wavelengths: atmosphere, surface material characterization, biological processes.

Figure 18. Imaging approach for the SGL telescope.

Imaging with the SGL is done on a pixel-by-pixel basis (Figure 17). The spacecraft must conduct a raster scan, while moving in the image plane (Figure 18). Each pointing corresponds to an Einstein ring for the entire planet that is composed from individual Einstein rings formed by a large number of surface pixels. Between the adjacent telescope pointings (i.e., pixels) the impact parameter changes, which corresponds to a slightly different instantaneous optical axis. For each such axis, the light in the Einstein ring would be dominated by a particular surface pixel, with contributions from a large number of adjacent surface pixels on the planet. Therefore, depending on the instantaneous optical axis, the total ring will have a brightness corresponding to a particular orientation of the planet-Sun-telescope system.

To build a $10^3 \times 10^3$ pixel image, we will sample the image in a pixel-by-pixel fashion, while moving in the image plane with 1000 steps over 1.3 km with resolution of 1.3 m. This can be achieved relying on a combination of inertial navigation and 3 laser beacon spacecraft placed in 1 AU solar orbit whose orbital plane is co-planar to the image plane. (For Proxima b, one would just have to point the telescope at a spot which the planet would re-visit each 11-day period and continue along that trajectory, accounting for the proper motion).

Light contamination from the parent star is a major problem for all modern planet-hunting concepts. Due to the SGL's ultra-high angular resolution (~0.1 nas) and a very narrow FOV, the parent star is completely resolved from the planet with its light amplified $\sim 10^4$ km away from the optical axis, making the parent star contamination issue negligible in the scenario involving the SGL.

To evaluate the imaging performance of the SGL, we consider an exo-Earth at a distance of 100 light years away and account for zodiacal background, the solar corona brightness, spacecraft jitter, and other realistic losses, etc. We assume that the instrument will have coronagraphic suppression of $10^6$. Using these assumptions, we estimate that a 1-m telescope, operating at 650 AU on the focal line of the SGL, would reach a SNR of 7 in ~7 weeks of integration time. Such a performance is sufficient to image this target with a resolution of $10^3 \times 10^3$ pixels.

This work allowed us to consider two approaches for image deconvolution, relying on the rotational and the direct deconvolution methods.

### *2.3.1 Rotational deconvolution*

We performed a simulation of the rotational image reconstruction. There is more than one way to create a multipixel image of a distant exoplanet. In this phase of the study, we examined a technique that is sometimes called rotational tomography. The basic idea is that with the highly aberrated image of the exo-Earth at the SGL focus, we detect light from the whole planet. If we block out light from a part of the planet, say a 50 × 50 km region, it will result in a very slight dimming of the total flux. The difference between these two numbers is the flux from the 50 × 50 km region that was blocked. The planetary rotation causes different parts of the planet to rotate into view over a day. This approach allows one to use a 1-day light curve and perform longitudinal deconvolution.

We use the images of the Earth taken by the EPIC camera on the NASA's Deep Space Climate Observatory[3] which is at the Sun-Earth L1 Lagrangian point. The raw data is being released as near-real time images of Earth through the EPIC instrument's website[4]. For our purposes, we ignored the fact that clouds come and go, moving in the Earth's atmosphere. The input data is used to produce an albedo map parameterized by longitudes and latitudes. The half of the planet that is visible at any given time is integrated to produce a light curve that represents the variation of the total flux, as seen by the SGL (Figure 19).

This type of deconvolution requires a very high photometric SNR. The SNR loss in doing this type of deconvolution is roughly proportional to the number of pixels. Also, this produces not an image of the planet's surface but an albedo map. If we have 100 pixels in latitude and 300 in longitude, the 30,000-pixel albedo map will have roughly 30,000 hits on the photometric SNR. Even a 1 km filled aperture telescope would not be enough for this purpose. Fortunately, the very large effective collecting area of the SGL allows for such a SNR. We plan to continue this work in Phase II.

[3] Deep Space Climate Observatory (or DSCOVR), https://en.wikipedia.org/wiki/Deep_Space_Climate_Observatory
[4] Website for Earth Polychromatic Imaging Camera (EPIC), https://epic.gsfc.nasa.gov/

This light curve can be inverted by using a straightforward pseudo-inverse procedure in MATLAB to produce a longitudinal map of the planet. Assuming that the planet's spin axis is not exactly aligned with its orbital axis, over the course of a year the illumination over latitude also changes.

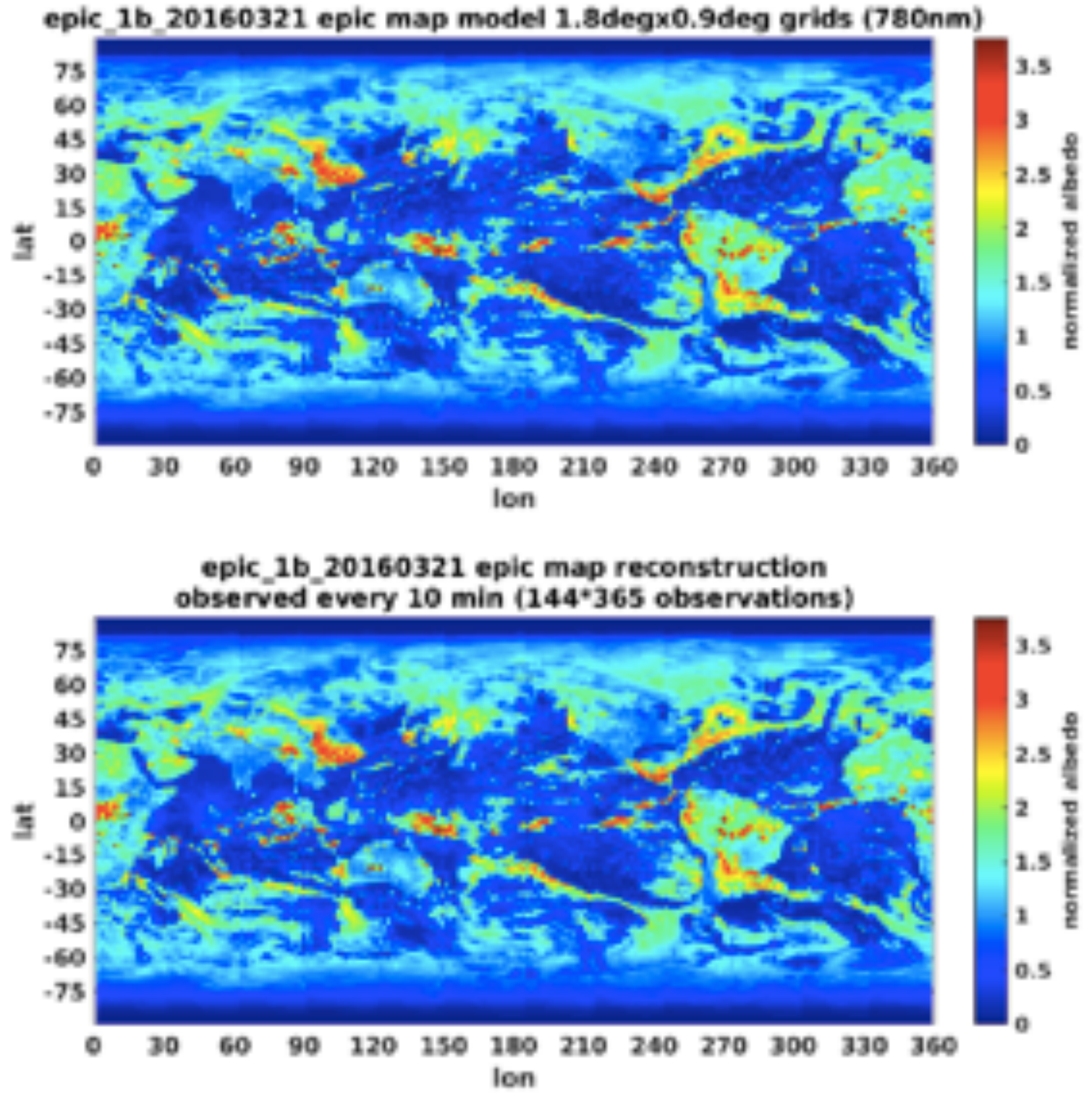

Figure 19. Rotational deconvolution: Top is the input map, which is the albedo map of the Earth used to generate light curves. The bottom is the albedo map calculated by "inverting" the light curves back to a 2D image.

- Significant potential for high-resolution spectroscopy:
  - Spectroscopic signal is high: SNR~$10^6$ in 1 sec (broadband);
  - Splitting this signal in ~$10^6$ bands would yield high-resolution spectra.
- Although very powerful, the Sun is not a very good lens:
  - Magnified images will be highly blurred, with any given pixel containing light reflected from adjacent regions on the surface of the exoplanet.
  - Would require correction with modern image reconstruction techniques:
    - Planetary rotation would provide periodic changes. Rotational deconvolution (aka tomography): ~250-300 pixel images in ~1 year;
    - Direct deconvolution: SNR reduction because of blurring, leading to longer integration times to reach ~800 pixel (study is ongoing).
- Observing on the background of the solar corona:
  - Corona pushes us to higher heliocentric distances ~650-850 AU;
  - Required chronographic performance $10^{-6}$ (WFIRST needs $10^{-9}$);
    - Internal coronagraph that has ~$2\times10^{-7}$ attenuation, 15% throughput, but pushes to larger apertures 1.5-2.2m;
    - External coronagraph (i.e., starshade) allows for smaller aperture(s).

Figure 20. Exploring optical and spectroscopic properties of the SGL.

Given the enormous amplification of the SGL, we will consider doing spectroscopic research of the exoplanet, even spectro-polarimetry (Figure 20). It will not just be an image, but potentially a spectrally resolved image over a broad range of wavelengths, providing a powerful diagnostic for the atmosphere, surface material characterization, and biological processes on an exo-Earth.

### *2.3.2 Deconvolving the exoplanet image from measurements of the Einstein ring*

Creating a megapixel image will require at least one million separate measurements. For a typical photograph, each detector pixel within the camera is performing a separate measurement.

This is not the case for the SGL. Only the pixels in the telescope detector that image the Einstein ring are measuring the exoplanet, and the Einstein ring contains combined information from the entire exoplanet surface, due both to the blur of the SGL and also to the relative distribution of different regions of the exoplanet to different azimuths of the ring. Many suggestions have been made about how to exploit features like azimuthal information to better resolve the exoplanet.

We adopt here a much simpler measurement scheme, in order to enable a straightforward measurement and deconvolution method that we can then use to estimate required signal-to-noise ratios needed to estimate integration times with the SGL.

We assume that the exoplanet is always in phase, that it is not rotating, that its features never vary with time (no clouds passing, no seasons), and that the telescope can maneuver around the exoplanet image at the SGL at will. We further assume that the corona image at the telescope can be predicted by independent means (that is, without using SGL telescope time), and subtracted from the Einstein ring image with shot noise limited precision. This will be a key tradeoff for our Phase II study: given the planet's rotation, how quickly can we collect the pixels and in what pattern would this be most efficient?

With these assumptions, we have a simple concept to form a megapixel image of the exoplanet:

1) Divide the SGL image plane into a $10^3 \times 10^3$ grid that just contains the entire exoplanet in the 'directly imaged' sense.
2) Position the telescope in each pixel for time $T$ and measure the total power at that pixel in the region of the telescope detector containing the Einstein ring, $10^6$ measurements in all.
3) Subtract the independently determined coronal power for each pixel to get the Einstein ring contribution at that pixel.
4) Multiply the vector of image pixels by a deconvolution matrix to obtain a vector of object pixels at the exoplanet.

Step 4 requires elaboration. A given pixel in the image plane will receive light from every pixel in the object plane, due to the blur. We represent the source object by a vector $\bar{S}$ of one million elements, one for each pixel, and the resulting image by a similar vector $\bar{R}$, where the source pixel $S_i$ is directly imaged by $R_i$ for all $i$. We mathematically represent the convolution caused by the blur by the square matrix $\bar{\bar{B}}$, which has $10^6 \times 10^6$ elements; each element $B_{ij}$ of $\bar{\bar{B}}$ contains the gain at the image pixel $i$ for the source pixel $j$, which can be calculated from the PSF. The coronal signal and other sources of measurement noise at each pixel is represented by the vector $\bar{C}$, and the measurement is given by the vector $\bar{M} = \bar{R} + \bar{C} = \bar{\bar{B}} \cdot \bar{S} + \bar{C}$.

In principle, if the convolution matrix $\bar{\bar{B}}$ and the measurement noise $\bar{C}$ are both known, this equation can be solved for $\bar{S}$ by inverting $\bar{\bar{B}}$. In reality, only the statistical properties of $\bar{C}$ are known, and $\bar{\bar{B}}$ is too large for efficient matrix inversion. However, alternative algorithms exist to estimate $\bar{S}$ accurately and efficiently, e.g., the iterative Richardson-Lucy algorithm, applicable when the noise is Poisson-distributed, or algorithms relying on the nearly diagonal nature of $\bar{\bar{B}}$.

If we wish to determine the elements of $\bar{S}$ with a given $\mathrm{SNR}_S$, assuming all $N$ source pixels contribute equally to a given image pixel, image pixels must be measured with $\mathrm{SNR}_S = \mathrm{SNR}_R^N$.

We generated examples of $\bar{\bar{B}}$ of much lower dimension $N$, and found that: 1) each row of $\bar{\bar{B}}^{-1}$ consists of a single positive element on the diagonal, with negative elements everywhere else, and

2) the sum of the negative elements in each row is about $(N-1)/N$ times the single positive element on the diagonal. In other words, every measurement of the image at some pixel will be $N$ times larger than the 'unblurred' values, and to get the unblurred value we must subtract contributions from all other pixels to about $1/N$ precision.

Error sources include the proper motion of the parent star, orbital motion of the planet around it, spacecraft logistics in sampling the image plane, wave front sensing, etc. We evaluated the contributions of these error sources on imaging.

The effect of blur, and thus the total integration time can be greatly reduced if the exoplanet is observed in a crescent phase and we know which part of the exoplanet is illuminated. For instance, when only 10% of the exoplanet is illuminated, this means only 10% of the total number of pixels. The blur per pixel is also reduced by a factor of ten. Therefore, the required integration time is reduced by a factor of 100. Repeating this process ten times while different areas of the planet are illuminated, we obtain an image of the entire exoplanet at the desired resolution in one tenth the time. This improvement is entirely due to reducing the effects of blur. Even thinner crescent phases would allow even more rapid measurement, but such extreme crescent phases occur rarely, and it may be necessary to consider blur introduced by the exoplanet's much brighter host star, which will be at small angular separation.

Based on this approach, we demonstrated that with a 1-m telescope, we would need ~2 years to build an image of 500 × 500 pixels. We think that any data collection time less than 5 years is acceptable. The integration time is proportional to the number *N* of desired pixels and the fourth power of the distance to the target: $T \propto ND^4$. Another scaling law exists, which is related to the diameter of the telescope, *d*. A telescope with twice the aperture will collect four times more photons. In addition, its diffraction pattern will be twice narrower and, thus, it will collect twice fewer corona photons. As a result, the integration time will scale as $T \propto 1/d^3$. Therefore, an image with 1000 × 1000 pixels may be produced in ~5 years with a 2-m telescope. The rotational motion of the crescent exo-Earth is also worth exploring in a detailed simulation of rotational deconvolution. Another factor that influences the integration time is the heliocentric distance and the corresponding size of the Einstein ring. However, the two factors that reduce integration times by one or more orders of magnitude are the pixel resolution and the telescope aperture.

As a result, we were able to identify a photon/fundamental limit to the deconvolution. We were able to find a number of parameters that would allow us to do a tradeoff study needed to identify the most optimal mission design. There are still may a number of very difficult practical problems that are yet to be solved (and perhaps yet to be identified.) However, we already identified a set of key driving parameters, including a) heliocentric distance, b) aperture of the telescope, c) integration time, d) detector sensitivity and its type, e) coronagraph and starshade performance. We will continue this study during Phase II.

### *2.4 Heliocentric ranges for an SGL mission*

By using light rays with larger impact parameters, *b*, we can limit the noise from the solar corona brightness, simplifying coronagraph design. At 550 AU, the Sun subtends ~3.5". At $\lambda = 0.6$ µm, the diffraction-limited size of a 1-m telescope is $\lambda/d \sim 0.1$" or 35 times smaller than the diameter of the Sun. A $3\lambda/D = 0.3$" coronagraph is ~1/5 of the solar diameter. Thus, there is no need to go to large distances to compensate for solar corona brightness. The value of $b = 1.1R_\odot$ corresponds to the heliocentric distance of ~660 AU. From that distance, the Sun itself subtends 2.9", while the circle with radius of $b = 1.1R_\odot$ is 3.2". If we could stop at 660 AU, all we would need to do is to move in the image plane to build an image of the exo-planet.

Hoverer, reaching these distances in ~20 years implies moving at a radial velocity of >20 AU/year, which makes stopping impractical. Instead, the spacecraft will proceed outwards while it continues the imaging campaign. An impact parameter of $b = 1.25R_\odot$ corresponds to the distance of $F = 854$ AU. From that distance, the Sun subtends 2.24", while the circle with radius of $b = 1.25R_\odot$ subtends 2.81". Although the lens could be used all the way to ~1,500 AU, the heliocentric region of interest for our mission would be the range 660–850 AU. To minimize the impact of the solar plasma on the instrument performance, the range may be extended a little, to 700-950 AU. This choice will be explored further when we conduct a realistic simulation that would include direct image deconvolution. This work allows us to explore some mission design parameters vs. image quality and related integration time in a realistic setting.

### *2.5 The a priori properties of the target*

From the beginning we treat the SGL telescope as a single-target instrument. That is to say that the target must be of high enough value so that we may embark on development of the mission to the SGL. As was addressed in our KISS Final Report (Stone et al., 2015), there may be other important science objectives, such as a flyby of a distant protoplanet in the Kuiper Belt, a number of interesting investigations relying on the parallax science en route to the focal region of the SGL, search for and investigations of exoplanets in our stellar neighborhood, a number of high precision tests of fundamental and gravitational physics, study of the solar system's dust background, ultra-deep field investigations of the cosmological background from the SGL and others. However, the mission must remain strongly focused on its primary objective: direct imaging and spectroscopic investigations of a distant exoplanet. Therefore, the target must be well-justified.

Towards that goal, we assume that given that small planets are expected to be ubiquitous (Fressin et al. 2013), in the coming decade we anticipate learning about Earth-like exoplanets with atmospheres, free oxygen, water, etc. Evaluating what we may already know about the exoplanet (e.g., rotation period, prevalence of clouds) will be important to establish mission requirements, optimizing the reconstruction of a spatially resolved image and motivating precursor projects.

We witness the intensified progress in exoplanet discoveries: Kepler has detected a plethora of potential Earth-like exoplanets, placing the possibility that another Earth-like world exists into the public consciousness. Follow-ups on Kepler candidates with other current exoplanet characterization technologies yields unresolved images at low spectral resolution (typically $R < 100$). The next steps in the remote exploration of exoplanets include TESS (2017), which will extend Kepler's work by performing an all-sky survey to identify additional exoplanet candidates, including Earth-like planets; JWST (2018), which will be used for targeted follow-up on candidate planets; and missions in formulation, such as the Exo-C (2015), Exo-S (2015) and LUVOIR (2015) concepts.

JWST will likely spend months of observing time evaluating atmospheric properties (possibly bi-omarkers) on a limited number of targets (possibly only one). Even with JWST, there is a possibility for only a marginal detection (Deming et al. 2009); the same is true for the 30-m ground-based telescopes. Without the large number of Kepler detections, to date ~62% of known exoplanets are within 100 pc, and almost all (~80%) of the Earth-like (i.e., super-Earth) planets are within 100 pc. This gives us a great starting point for a candidate target list for the mission to the SGL.

The occurrence rate of Earth-sized terrestrial planets in the habitable zones (HZs) of Sun-like (FGK) stars remains a much-debated quantity. Only a handful of such planets have been discovered (e.g. Torres et al., 2015). Current estimates range from 2% (Foreman-Mackey et al., 2014) to 22% (Catanzarite & Shao 2011, Petigura et al., 2013). The Simbad database lists 8,589 F stars, 5,309 G stars, and 1688 K stars within 30 pc. Taking even the lowest estimates, we can expect to detect at least one terrestrial planet in the HZ of a star within 30 pc in the near future. Once such a planet is discovered, significant observational resources will be devoted to characterize it.

- We want to image Earth 2.0, around a G star, which is not transiting:
  – Once habitability is confirmed ("big TPF" for spectra), the next step is to image it.
- We will rely on astrometry, RV, spectroscopy, and direct imaging to obtain:
  – orbital ephemeris: to ~mas accuracy and precision;
  – rotation: from temporal monitoring of the spectroscopy;
  – atmosphere: temperature, structure, chemical composition, and albedo, from non-spatially-resolved spectroscopy;
  – understanding of cloud & surface properties from Doppler imaging.
- This information will help us to point the s/c:
  – Time to reach 550 AU ~10 years, enough to observe the parent star's location ~100 times with 1 $\mu$as precision, so that its position would be known to 0.1 $\mu$as;
  – The parent star's position would be known to ~45 km at a distance of 30 pc;
  – Orbital period to <1% $\Rightarrow$ the semi-major axis is known to ~0.7% (~1 million km);
  – If face-on, the radial distance to ~1 million km, with tangential error ~6 larger;
  – Earth's diameter is 13,000 km, so we will search the (80 × 500) grid on the sky;
  – Once SGLFM detects the planet $\Rightarrow$ scan a smaller area to define the "edges".

Figure 21. The a priori properties of the target.

Most likely, we will want to image Earth 2.0, around a G star which is not transiting (Figure 21). An SGL mission could follow a "big TPF" that observes an exo-Earth around a G star and measures its spectra. We should be very confident that the selected target is habitable. A spacecraft at the SGL would be the next major step, possibly the biggest step in the 21st century for exoplanet exploration. If the planetary atmosphere contains oxygen and, possibly, signs of life, the next step would be to launch the mission to the focus of the SGL to image this planet. The planet's orbit would have to be measured in 3D, using either astrometry and/or RV measurements combined with direct imaging. With luck, it will be inclined so that it transits, providing a radius. These measurements would allow obtaining that information and point the spacecraft.

Once we know of a terrestrial HZ planet so close to our own, we posit that significant resources will be devoted to characterizing the planet and its system using the above techniques. The knowledge we gain from this will include: i) orbital ephemeris, to at least milliarcsecond accuracy and precision, ii) detailed knowledge of the atmosphere, including temperature, structure, chemical composition, and albedo, all inferred from non-spatially-resolved spectroscopy; iii) estimates of rotation rate, gained from temporal monitoring of the spectroscopy, and iv) some understanding of cloud and surface properties from Doppler imaging (Crossfield et al., 2014).

Therefore, we assume that spectroscopic biomarkers will have been found by the time we launch the mission to the SGL. This discovery must be confirmed by following a disciplined process. Figure 22 shows details regarding the logic of the discovery of life and its conformation. In fact, it is likely that life will be found by traditional techniques—transit or direct spectroscopy—either with JWST, one of the giant segmented mirror ground-based telescopes, or concepts like HabEx and/or LUVOIR. The mission to the SGL will follow these efforts. What the mission to the SGL can do, however, is spatially resolved spectroscopy and identification of where these biomarkers come from. They will likely be associated with biomasses (e.g. forests, swarms of plankton, etc.) that can only be spatially resolved by the SGL telescope. SGL is uniquely capable to conduct spectroscopic observations of an exoplanet. With a very strong photometric gain, the lens allows for high resolution spectroscopy that may lead to unambiguous identification of life on an alien planet. We plan to address the spectroscopic capabilities of the SGL during our Phase II effort.

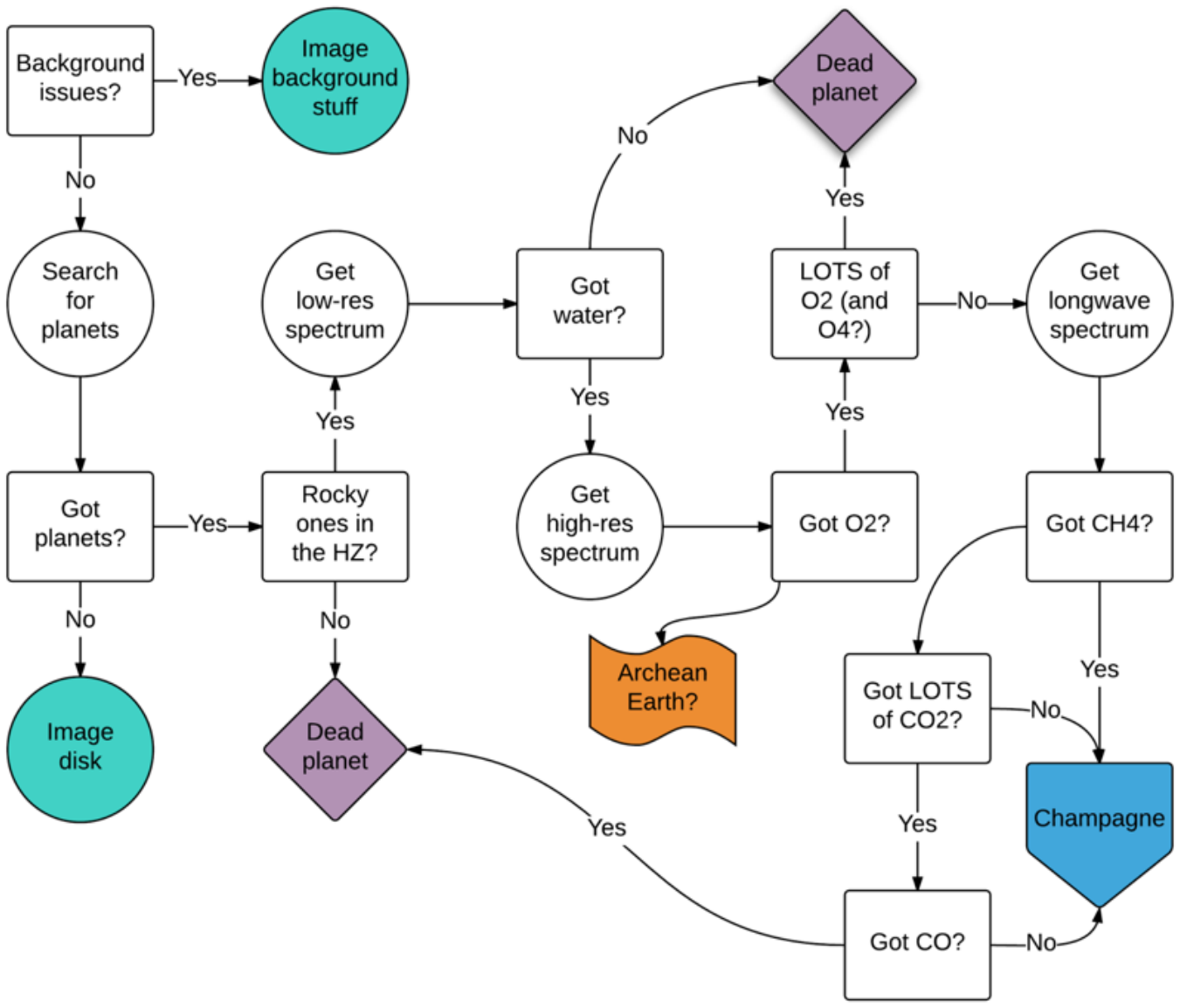

Figure 22. A life-searching process (from S. Domagal-Goldman.) What is missing here the ability to detect lights on the night side – one of the great benefits of the SGL.

Therefore, we think that the mission to the SGL would begin after the discovery of an exo-Earth, and there would be ~10 years of "cruise" before the spacecraft would reach ~550 AU. During those 10 years, the parent star's location would have been observed with microarcsecond (μas) precision at least a hundred times, so that its position would be known at the 0.1 μas level. The parent star's position would be known within ~45 km at 30 pc. The orbital period of the planet would be known to <1% meaning that the semi-major axis is known to ~0.7% or ~1 million km. If the planet is in a face-on orbit, we will know the radial distance to ~1 million km, but the error bar in the tangential direction will be a factor of ~6 larger. The diameter of the Earth is 13,000 km, so that the area on the sky we must search is an (80 × 500) grid. Once the SGL telescope detects the planet, it would scan a much smaller area to define the "edges" of the planet. Astrometry of the star when the planet is discovered would have measured its mass; that and its size give us the density of the planet.

This bring us to the main science objective of our mission: what information do we have to collect to unambiguously detect and study life on another planet? To answer this question, we would first have to clarify are we looking for Earth 2.0 and life like ours. The concept of a "habitable zone" around a star is predicated upon this concept. That means that the planet-star distance is such that an equilibrium model of a planet with an atmosphere but without a strong greenhouse gas would yield liquid water on the surface of that planet. Within that constraint, Venus is too close and Mars is too far away. Hence the idea of the Goldilocks Zone. Within that definition, Jupiter and Saturn are clearly outside the habitable zone, yet they have liquid water clouds and icy satellites that have liquid water interiors and are plausible abodes for life. But they are not Earth 2.0.

If we are only interested in Earth 2.0 that can be detected remotely, atmospheric gases are still the best option. However, planets with Earth's volatile inventory can evolve along very different courses as planetary evolution models are showing. There are geochemical non-life scenarios that result in $O_2$ in the atmosphere, and even 60 bars of $O_2$, but no liquid water because of $H_2$ escape. Hence these models are desiccated.

Perhaps, we should look for a mixture of gases that are clearly in disequilibrium with each other or are quickly photochemically destroyed but are nevertheless there. In the case of Earth, the presence of both $O_2$ and $CH_4$ would be a strong indicator of life. Life essentially is creating disequilibria and that can show up in the planet's atmosphere. But only in the case of planets with an atmosphere where the biosphere has had sufficient influence on the planet to alter the composition of the atmosphere. Note that it took at least two billion years to accomplish that on Earth.

To address this possibility, we studied spectroscopy and photometry with the SGL. The broadband photometric SNR is ~$10^3$ in 1 sec of integration for a 32 mag object. At higher spectral resolution, the needed integration time would be higher. But it is highly likely that the SGL mission will be able to detect both methane and oxygen and likely many other molecules in the atmosphere of an exo-Earth. The spectroscopic SNR of the SGL telescope is very respectable.

The same mission may also be able to image of all the planets orbiting that star which also conducting spectroscopic investigations of their atmospheres. This is an attractive feature of the mission, as most of the discovered planetary systems to day indicate that planets come in families. Therefore, the imaging mission to the SGL will have multiple targets to study.

These considerations of the optical properties of the SGL, impact from the solar corona brightness, instrument design, various types of deconvolution processes available, as well as the properties of a potential target helped us establish preliminary science and mission requirements.

# 3 CONCEPT FOR A MISSION TO THE SOLAR GRAVITY LENS FOCAL AREA

The SGL offers a unique means for imaging exoplanets and determining their habitability. Theoretical considerations (Turyshev & Toth, 2017) are promising, both for getting there and for capturing high-resolution images and spectra of potentially habitable exoplanet. The mission concept has the potential of being the most (and perhaps only) practical and cost-effective way of obtaining kilometer-scale resolution of a habitable exoplanet, discovering and studying life on other worlds.

A complete set of requirements to create such an image with the SGL is yet to be determined. Current knowledge of the optical properties of the SGL, preliminary instrumental designs and initial deconvolution analysis help us towards that goal. Exoplanet imaging requires several key technologies that are challenging. These include the determination of an exoplanet's astrometric orbit at better than nas precision, and the motion and stabilization of the spacecraft over millions of pointings with limited power. These issues are hard but they are not unsolvable.

## *3.1 Study approach and design trades*

During Phase I of this effort, we studied several possible mission architectures, analyzing the tradeoff space available for this mission. Our objective was to explore the possibility of using a single spacecraft with enough on-board propulsion, navigation and pointing capabilities to move pixel-by-pixel in the image plane. An alternative architecture would rely on a pair of spacecraft connected with a boom or tether of variable length (see a video depiction done in (DeLuca, 2017)). One could also fly a swarm of small and maneuverable spacecraft with a mother craft on the focal line. This approach allows probing the spatial structure of the caustic as a function of time and has the potential of allowing sampling, modeling, and subsequent (on-board) removal of imaging systematic errors due to possible radial, azimuthal, and temporal departures from the idealized caustic structure. We explored the role of swarms in reducing navigational and maneuvering requirements for individual spacecraft due to proper motion of the exoplanet, its orbital motion, and rotation.

With this approach, we studied several topics related to mission design, specifically: i) flight system and science requirements; ii) key mission, system, and operations concepts and technology drivers; iii) description of mission and small craft concepts with navigation and system design to reach and operate at the SGL; and iv) instruments and systems for the SGL spacecraft, including power, communication, navigation, propulsion, pointing, and coronagraph.

### *3.2 Elements of the SGL mission concept design*

Previous deep space mission concept studies assessed the feasibility of reaching beyond the solar system (e.g., Etchegaray 1987, Nock 1987, Liewer et al. 2000, see ISP 2015). These include several NIAC studies (McNutt et al. 2003, Nosanov et al. 2013), most recently (Cesarone et al 2014; Stone et al. 2014). West (1999) reported on a study of a mission to 550 AU with the objective of using RF technologies in conjunction with the SGL.

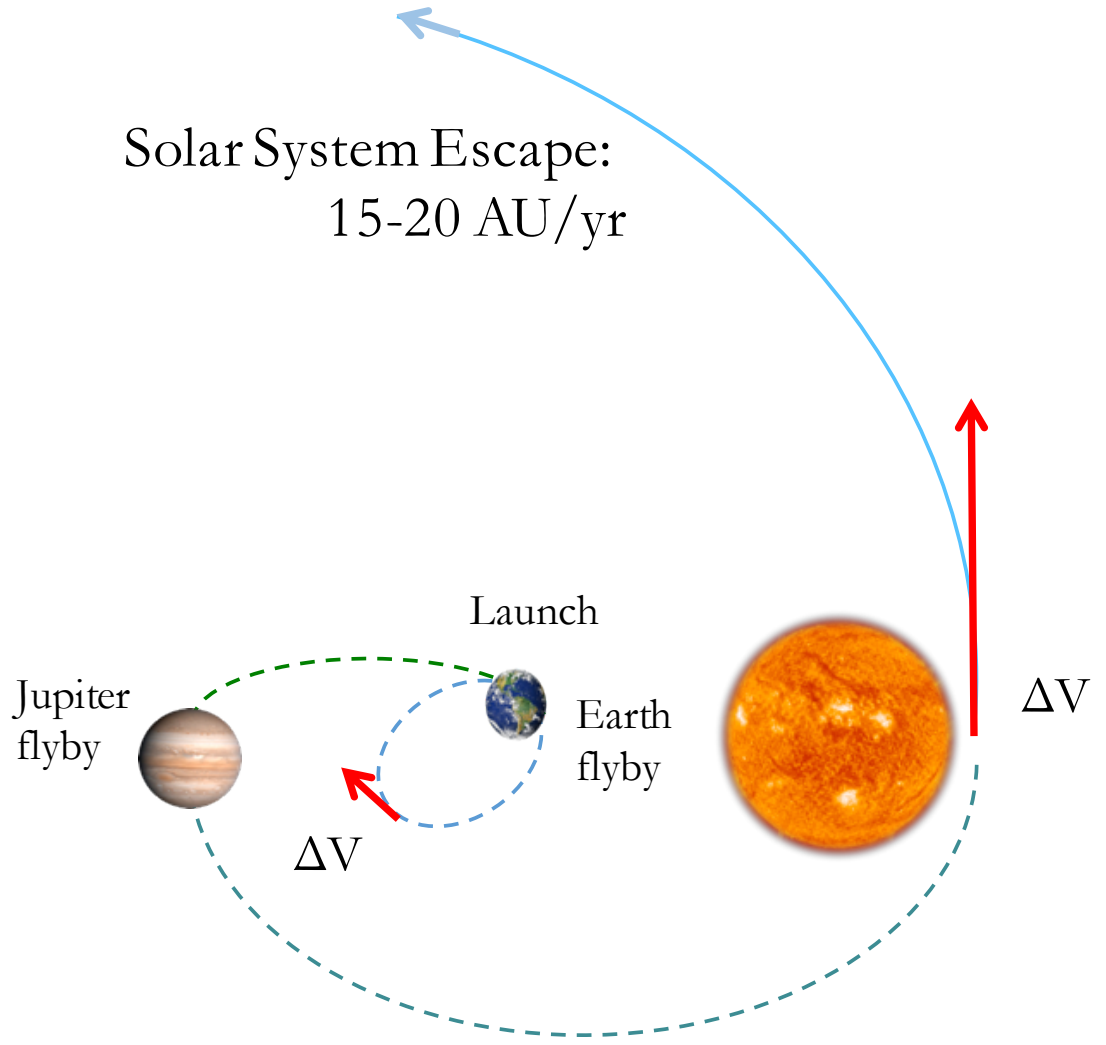

Figure 23. Baseline mission concept for reaching 650AU in ~40 years with conventional chemical propulsion technology.

That study emphasized the need for new technologies including advanced propulsion, lightweight telescopes, membrane mirrors, inflatable/rigidizeable structures, and novel coronagraphic techniques. Some of these are technologies either already available or else being developed (Stone et al. 2015). Others, including the high-performance propulsion required to achieve the study's objectives of a 3–10 year transit to the 550 AU point, remain beyond the state-of-the-art for near-term mission panning, including solar sails with diameters exceeding 1 km and nuclear powered ion engines with $I_{sp}$ exceeding (15–20) × $10^3$. We will benefit from these earlier studies of deep space exploration in the mission design, focusing on the SGL-specific challenges.

To achieve distances beyond 550 AU with Voyager-class mission durations, our baseline mission concept would employ a Jupiter flyby followed by a low-perihelion Oberth escape maneuver. The KISS 2014 study (Stone et al. 2015) considered chemical escape maneuvers using thermal shielding technology at ranges as low as $3R_\odot$. Alternatively, solar sails could allow high escape velocities with perihelia of $20R_\odot$ (0.1 AU), but would require sail area-to-mass ratios larger than the current state of the art (Friedman & Garber 2014). We will trade these two propulsion approaches. Either approach could achieve an escape velocity of 15–20 AU/year and reach the SGL in 30–40 years (Figure 23).

Recent results in nanosatellites and small spacecraft development have suggested a practical mission concept to reach large distances previously beyond our capabilities. The new capabilities would allow a spacecraft to fly along the focal line to image a preselected exoplanet at high resolution, which is impossible to achieve in any other way. Future study topics should include analysis of nanosatellite technology for very deep space flight and instruments including communication, power, stability, control, raster scan in the image plane, mission design concepts and payload possibilities.

Early in the study, we identified the most important challenges related to i) the choice of propulsion technology (how to get the large heliocentric ranges and how to operate at the SGL), ii) the size of the spacecraft (either a large and capable spacecraft or a swarm of smaller nanosatellites), iii) the concept of operations while at the SGL focal region, and iv) communications (how to transmit data back to Earth). There are many other challenges, but the ones listed are mission-enabling and must be dealt with first.

A number of modern technologies may enable a meaningful step to venture significantly beyond our solar system, were identified in the KISS study (Stone et al., 2014, 2015) and in a companion JPL study (Cesarone et al., 2014). Voyager 1 took 37 years to travel 132 AU from the Sun and is currently moving at a heliocentric velocity of ~17 km/s. An order of magnitude increase in this speed may be possible with near-term technologies. Large area-to-mass ratios for the solar sail require consideration of small spacecraft (e.g., nanosatellites), a promising enabling technology. Instrument capabilities for small spacecraft will be considered. In addition to solar sails and spacecraft requirements, key technologies of communications and power will be analyzed. The work on the technologies that enable reaching, and communicating from, large heliocentric distances is ongoing. In fact, there are already realistic mission concepts capable of reaching 250+ AU.

To achieve heliocentric distances of 550–1,000 AU in practical mission times of 25–35 years, we consider the following technologies:

- Optical communication allows for low mass, low volume, and low power, which are critical when considering a mission to deep space.
- Modern small satellites were already considered for many solar system applications. They also provide the necessary condition to implement low-power electric propulsion.
- In terms of mission architecture, gravity assisted trajectories could benefit from close solar flybys to achieve a high solar system escape velocity.
- Use of lightweight radioisotope power generators could enable missions to deep space far from the Sun. Some of these were studied in previous NIAC studies (Nosanov et al., 2013).
- If a close solar flyby is chosen for the mission, one would have to rely on the advanced materials needed to protect spacecraft with solar (or electric) sails components in the immediate solar vicinity.
- Reliable, long-life operations on the way to and at the SGL would require advances in autonomy, control, and the design of adaptive systems.
- One must consider precision navigation of a spacecraft in the image plane. For pointing, one may need to investigate the use of laser beacons in a 1 AU heliocentric orbit that would enable communication, guidance and navigation for a spacecraft at the SGL.

Based on work conducted during Phase I of this study, we were able to identify the system requirements and architecture tradeoffs. This information aided our preliminary design concepts allowing us to assess key mission, system, and operations technology drivers.

Many of the advanced space technologies developed for precision experiments in space can be directly applied to the instrument and spacecraft design for a mission to the SGL. Such technological crosspollination allows other science disciplines to take advantages of space deployment opportunities, thereby stimulating the progress in many areas of space research.

### *3.3 Concept designs for an SGL mission*

Given the long travel times, a new kind of mission concept is needed to make an SGL mission possible. The main enabling aspect of such a mission would be the need to travel fast and survive longer. Whereas current technology cannot take us to another star anytime soon, we are poised today to target the SGL as an explicit destination for robotic exploration and science investigations as a first step on a long road ahead towards a potential Earth-like exoplanet.

One of the key challenges for this mission would be to explore innovative mission design techniques that would enable an SGL probe. Over the past two decades, there have been many reports in the literature that extol the merits of different ways for propelling robotic spacecraft as well as human spaceships towards the stars. These schemes include laser electric propulsion, nuclear fission, nuclear fusion, solar sail, laser sail, electric sail, microwave sail, magnetic sail, antimatter, and even extremely speculative concepts such as warp drives and zero-point energy propulsion. We took the approach of relying on conventional technologies only, which would allow us to develop a realistic mission design and then to conduct related technology tradeoffs.

Table 1. SGL mission architecture options: mission design tradeoffs required.

| System | Technology | Benefits, Costs, Requirements, Tradeoffs |
|---|---|---|
| Propulsion | Chemical | Big solid rocket burn very close to Sun; Massive shield;<br>Exit velocity limit ~18 AU/year |
| | Solar Sail | Lightest weight option, requires A/m beyond state of art<br>300x300 m sail with 100 kg sc -> 25 AU/y; with REP 30 Au/y |
| | Nuclear Electric | Expensive, heavy spacecraft, programmatic challenges |
| SC Size | 200-500 kg (conventional) | Allows state of the art design; Likely flagship mission development;<br>Probably required for chemical spacecraft;<br>Accommodates bigger optics. |
| | <100 kg (smallsat) | Will lower cost, consistent with multiple s/c. May be mission enabling<br>Consistent with "string of Pearls" architecture; Required for sail;<br>Requires new technology: e.g. low mass RTG; Imaging system, communications capabilities and system reliability TBD |
| ConOps | Single spacecraft | Straightforward, minimizes ops complexity |
| | Multiple spacecraft | 1) Creates more flexible mission design and data collection;<br>2) Consistent with "string of pearls" architecture and other relay or distributed data schemes; 3) Open architecture allows for incremented improvements in technology; 4) Provides important redundancy |
| **Comm** | Radio | May be enabled by use of sail as antenna |
| | **Optical** | **Likely better performance & implementation for given mass and power. Allows for optical ranging measurements** |

We considered the tradeoffs between a traditional telescope vs. a microsatellite system that opens up the possibility of sending multiple spacecraft (Table 1). We can devise an instrument that would rely on a swarm of small spacecraft, perhaps even launched together but each moving along a slightly different trajectory, parallel to the principal optical axis. Such an instrument would rely on the light collection capabilities enabled by a formation flying architecture.

Conceptually, many studies were made to investigate the science objectives and the technological feasibility of missions to deep space beyond the solar system, including several NIAC studies. Our work benefited from these earlier studies by allowing us to focus on the SGL-specific features. Clearly, a mission design to the focal region of the SGL presents a set of interesting challenges. Table 1 presents the summary of the mission architecture issues and the required mission tradeoffs. In this study, we looked at all of them at some level by considering three different classes of mission architectures: 1) a single and capable spacecraft, 2) a multi-spacecraft mission architecture, which we called a string-of-pearls, 3) a single small spacecraft relying on solar sail technology. By no means is this an exhaustive list of options for a prospective mission, but it allowed us to explore the entire tradeoff space which was kept loosely constrained.

### *3.3.1 A single probe*

Given the long mission durations of prospective interstellar probes, whether to the local interstellar medium (ISM), deep interstellar medium (ISM) or to the SGL and beyond, these missions are likely to have many common building block elements (Alkalai et al., 2017). For example, given that the ISM probes would be operating at large distances beyond the Sun, nuclear power (radioisotope generators – RTGs) would be required on all probes. For this study, advanced segmented modular RTGs (SMRTGs) were assumed. The SMRTGs are conceptual next generation of vacuum RTGs, capable of providing almost 5 times more power at their end-of-life (EOL) over the Mars Curiosity MMRTG and 2 times more power over the Cassini GPHS RTG. They would take advantage of skutterudite ($CoAs_3$) technology, which is already being matured for the proposed eMMRTG, use multi-foil insulation and aerogel encapsulation to achieve high efficiency and low degradation rate. Furthermore, their size can be optimized for a mission due to a segmented design.

The SGL mission concept was formulated to achieve an escape velocity of ~20 AU per year. This objective is particularly driving and requires a $\Delta V$ of >10 km/s at perihelion. Building upon the results from a recently published ISM trajectory design (Arora, et al. 2015) and KISS study (Stone et al., 2015) it became clear that achieving such a $\Delta V$ at perihelion would require moving beyond the traditional solid rocket motor (SRM) or bipropellant rocket engine technology. After a detailed propulsion trade study, two propulsion architectures stood out as viable candidates: 1) nuclear-thermal propulsion (NTP see (Larson et al., 1995)) and 2) solar-thermal propulsion.

Further system-level trade studies of the NTP technology let us to conclude that although the technology is being developed for possible future human Mars mission concepts, after factoring in the dry mass of the NTP stage, it cannot yet provide the required $\Delta V$. This resulted in the selection of an architecture based on solar-thermal propulsion (STP (Layman et al. 1998)) as the most viable option for this mission concept.

Next, we defined a notional baseline concept which was a result of a detailed mission design and multi-day JPL Team-X study. The objective of this Team-X study was to find a feasible point design which allows us to inform further development of the STP technology. The baseline launch stack for this mission concept consists of the following flight elements: 1). ISM Probe (~550 kg wet mass), 2). Perihelion maneuver stage (H2 tanks and a bipropellant system), 3). Solar-thermal propulsion system (including the heat shield, the heat exchanger and 12 engine nozzles).

The basic mission concept is optimized around achieving $\Delta V$ in excess of 11 km/s at the solar-perihelion using an STP system. The solar-thermal propulsion concept for solar-system escape application was first proposed in 2002 and relies on using hydrogen (cryocooled) as propellant, which is heated due to spacecraft's proximity to the Sun, using a heat exchanger, which also acts as part of a larger heatshield, designed to protect the spacecraft.

The mission concept requires a perihelion $\Delta V$ of ~11.2 km/s. The burn time is restricted to be less than 1.5 hrs. The probe uses an RTG-powered EP system providing an additional ~2.4 km/s of $\Delta V$. The escape velocity achieved is ~19.1 AU per year (or ~90.5 km/s). Given the high-$\Delta V$ requirements, the STP mission concept is very sensitive to the mass of the $LH_2$ tank, ISP, mass and support structure mass. Next, we summarize the main flight system components.

The STP system would consist of a double folded carbon-carbon heat shield (deployed after launch) with the middle panel also acting as a heat exchanger, 12 nozzles used for producing the required thrust, and a cryocooled $LH_2$ tank (Hastings et al., 2001). The heat shield, when deployed, is larger than the whole stack height, providing ample cooling and a shadow zone during the perihelion burn maneuver. The margined $LH_2$ tank mass is calculated using a tank mass factor of 39%.

The power requirement for the $LH_2$ cryocooler would be ~1.2 kW, which was estimated using a thermal model. The heat exchanger and the heat shield combination is of a multi-layered design and is estimated to be ~4 times as heavy as the one used in the original Thiokol (now Orbital ATK) study (Layman et al. 1998). The relatively large factor of safety accounts for details missed at this early stage in the design process.

During perihelion burn (at ~$3R_\odot$), $LH_2$ runs from the tank through the heat exchanger, where it heats up to 3,400 K and then passes through the 12 nozzles providing ~1,350s of ISP. The 3,400 K temperature is close to the carbon-carbon melting point of 3,800 K (Layman et al. 1998). This large ISP allows us to achieve the required mission design ΔV without excessive amounts of propellant.

Table 2. SGL probe concept mass allocation (Alkali et al., 2017).

| Sub-system | MEV, kg | Comments |
|---|---|---|
| Baseline power system (without SMRTG) | 47 | Ref. bus + batteries |
| Propulsion | 17 | Monopropellant |
| Communications | 30 | Iris level radio, 1-2 m deployable HGA |
| Mechanical | 177 | Light weight, multi-functional structures |
| Thermal 29 | 29 | RTG + RHU and Louvers |
| Attitude & Control | 34 | RWA, MIMU |
| C&DH | 15 | 3U JPL Avionics |
| Telescope | 50 | 0.5-1m Telescope |
| Propellant | 25 | Can be Xenon or can be extra mass for a bigger E-Sail |
| System Level Margin | 80 | According to JPL DP |
| 2x SMRTG | 52 | No margin needed for RTGs |
| **Total Allocation** | **556** | **Probe's wet mass** |

The conceptual probe, as designed during a Team-X session, is a monopropellant-based probe similar to New Horizons, with ~42 kg of instruments, totaling to a wet mass of ~561 kg. The data rate achieved from ~100 AU is ~200 bps. The probe would be powered using the advanced segmented modular RTGs (SMRTG3), capable of providing ~350 W after 15 years of life. Further design assumptions were made in the design, which lowered the probe wet mass to 430 kg. This allowed an addition of a RTG powered EP system with ~100 kg of Xenon. The total wet mass of the post Team-X REP probe is ~542 kg.

There is also a bipropellant system which is used to perform launch cleanup and trajectory correction maneuvers (TCMs) before the perihelion burn. The total launch stack also consists of a payload adapter and 4 extra SMRTGs connected to it. These RTGs are used for cryocooling the $LH_2$ and are dropped off just before the perihelion burn to maximize the available ΔV. The total stack mass allocation (including JPL 43% margin) is ~28,000 kg, which consists of 15,732 kg of $LH_2$, 620 kg of bi-propellant, and 11,278 kg of dry mass allocation. The dry mass allocation consists of 542 kg of probe mass (Table 2), ~1,842 kg mass for the heatshield and heat exchanger and rest allocated to support structure and 4 extra SMRTGs. The payload side adapter mass is estimated to be ~369 kg.

Alkalai et al. (2017) have considered a mission concept to the SGL focal region that should be able reach a distance of ~550–700 AU and deploy a 1 to 2 m size telescope for multi-pixel imagining of an exoplanet. The spacecraft would needs to reach the 600 AU range in < 40 years from launch. Factoring in the time required to build energy in the inner solar system for a type 2 trajectory, this results in an escape velocity of requirement > 20 AU/year. Hence, a baseline SGL mission concept would use advanced low-mass, low-power technologies, on-board autonomy and an advanced STP propulsion stage. Table 2 gives a flight system overview for a SGL probe.

As a result, we see that there already exists a feasible mission architecture that is capable of reaching the focal region of the SGL relying on modern-day technologies. The single-spacecraft mission concept has some obvious challenges (pretty much in every aspect of mission architecture and design), but it offers a baseline mission concept that may be used to evolve the design.

### *3.3.2 A solar sail mission concept*

We also considered a mission architecture based on solar sails. The key new technologies that enable consideration of such a mission are small satellites (spacecraft < 100 kg with power, communications, precision control and navigation, etc.) and solar sails. One interplanetary sail has already flown to the vicinity of Venus (JAXA's IKAROS) (van der Ha et. al., 2015) and another to a near-Earth asteroid is now being developed by NASA (NEA Scout) (McNutt et. al., 2014).

While conventional propulsion (chemical) could be used in principle with a large solid rocket motor flying very close to the Sun, even with optimistic assumptions the speed of such a probe is limited to about 17 AU per year (Stone et al., 2015). As described below, a 300 × 300 m solar sail, with a spacecraft mass of 100 kg could fly out of the solar system at ~25 AU per year, reaching the SGL in a less than 25 years of flight.

Heliocentric distances beyond 500 AU can be achieved in practical flight times with solar sails flying toward the Sun with a perihelion of 0.1–0.2 AU. Although the required spacecraft area-to-mass ratios are larger than the current state of the art, the requirements are consistent with those studied and considered in prior NASA and ESA studies. Other relevant propulsion technologies were also considered—an Oberth maneuver that was investigated in the KISS 2014 study (Stone et al., 2015), or electric propulsion—but solar (or possibly electric) sails appear to offer both the greatest performance potential and best near-term readiness.

In our recent study (Friedman & Turyshev, 2017), we assumed that a 200 × 200 m sail might achieve a solar system exit velocity of 25 AU/year. That could be achieved with a notional mass of 30 kg for the spacecraft bus; 13 kg for a radioisotope power system providing 100 W of electric power and possibly a small maneuvering capability; and a 1.6 kg sail with a density of ~0.04 g/m$^2$ (equivalent to 0.25-micron polyimide or a possible sail from carbon nanotubes). It remains to be determined if the baseline radioisotope power system can be smaller, or if it can contribute to the exit velocity with a propulsive boost.

The KISS 2014 study (Stone et al., 2015) considered pure chemical Oberth maneuvers near the Sun using Solar Probe Plus derived thermal shielding technology at ranges as low as $3R_\odot$. Alternatively, it was found that solar sails could allow high escape velocities with perihelia of $20R_\odot$ (0.1 AU), but would require sail area-to-mass ratios larger than the current state-of-the-art. Chemical propulsion is limited to an escape velocity of 15–16 AU per year, but solar sail trajectories may reach velocities ~25 to 30 AU per year. To go faster would require very advanced technologies, like a large nuclear reactor or space-based high-power laser propulsion.

Friedman & Garber (2014) first considered a mission to the SGL as an interstellar precursor. They studied solar sail requirements to reach exit velocities of over 20 AU per year. The results are summarized in Figure 24. Garber (2017) has extended this analysis to consider the area/mass requirements to reach an exit velocity of up to 40 AU/year. His result is given in Figure 25. Sail area to spacecraft mass ratios of 900 $m^2$/kg yield a speed of 25 AU/year, 30 AU/year requires an area-to-mass ratio of $A / m = 1{,}400$ $m^2$/kg and 40 AU/year requires $A / m = 2{,}550$ $m^2$/kg.

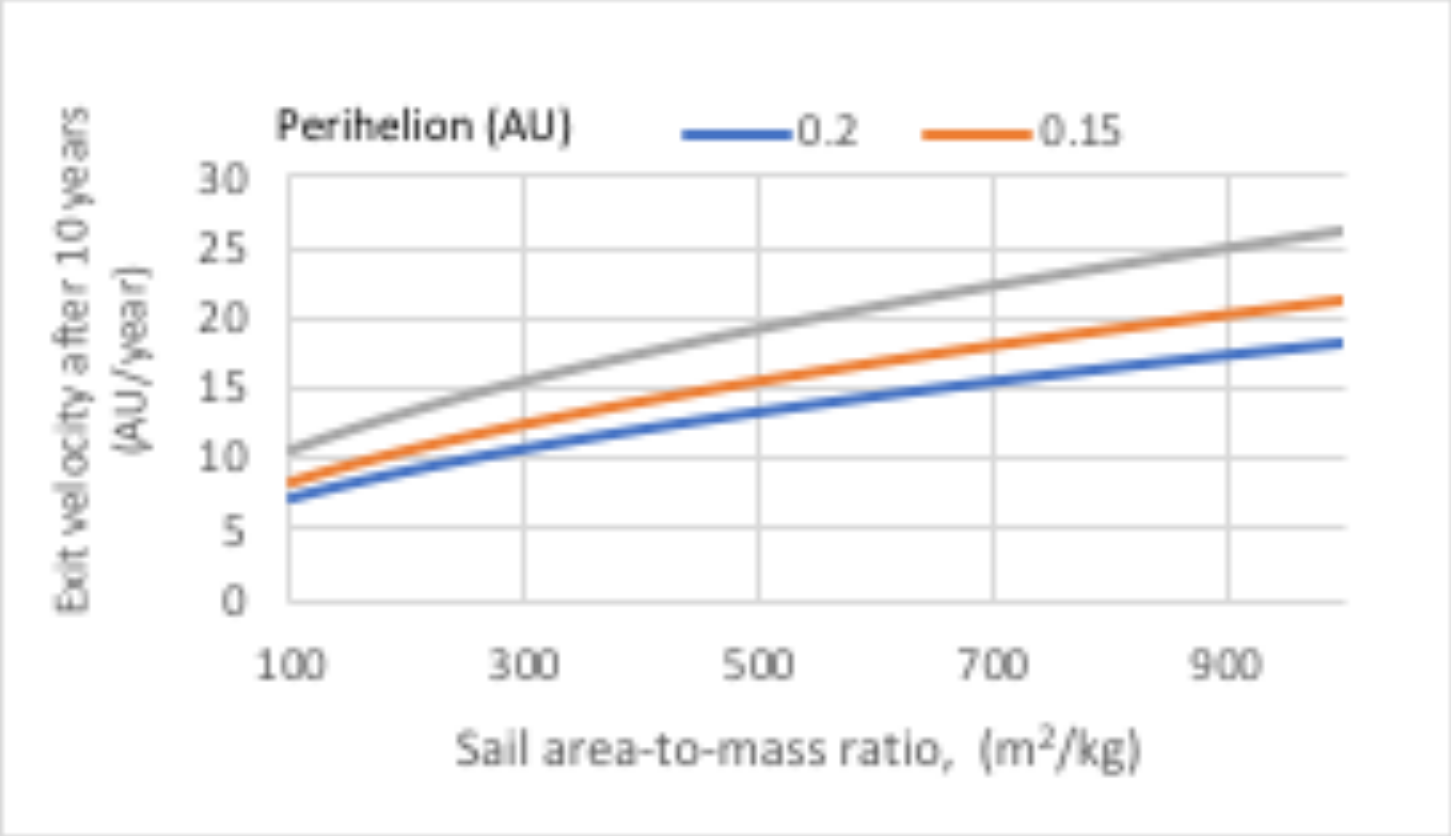

Figure 24. A solar sail spacecraft exit velocity from the solar system vs. sail area to spacecraft mass ratio and perihelion distance.

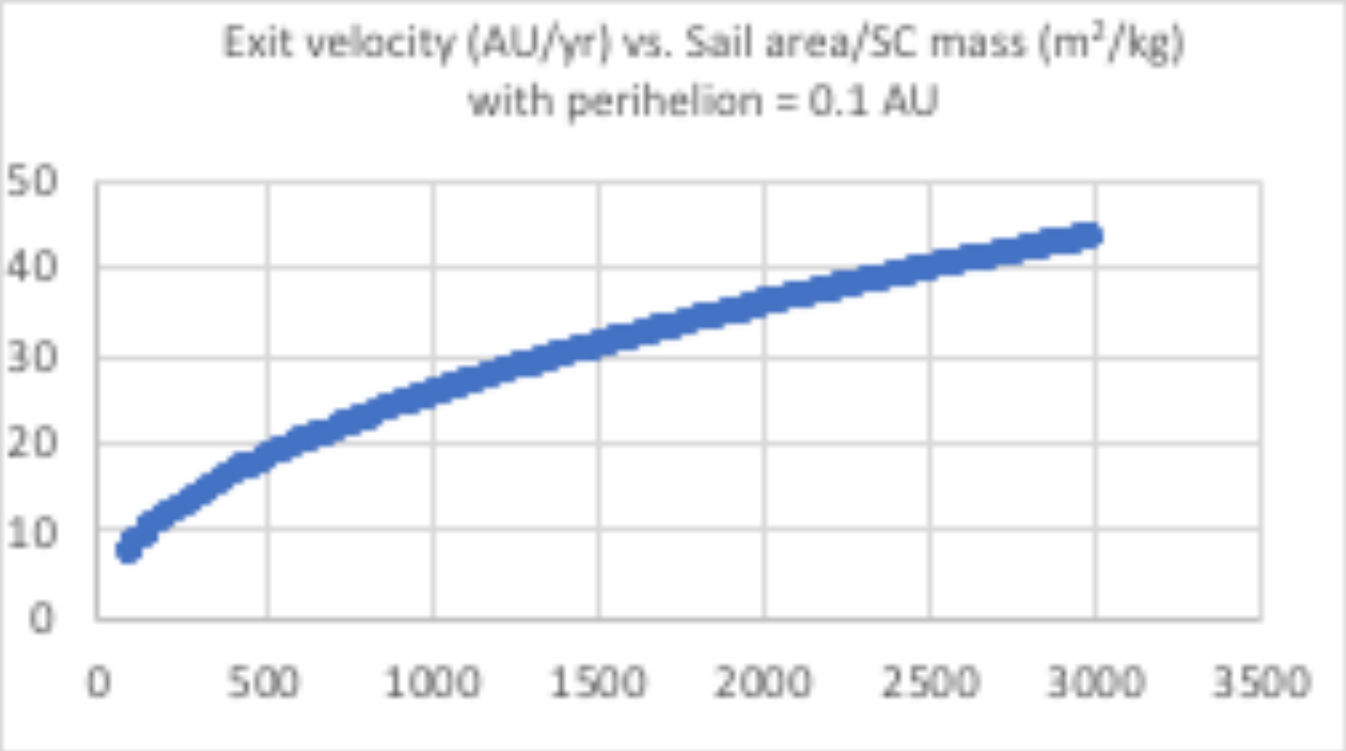

Figure 25. Exit velocity as a function of sail area/mass ratio.

Since any spacecraft needs power, presumably a small radioisotope generator, we consider that radioisotope electric power (REP) thrusters can provide an additional boost to the solar sail spacecraft as well as propulsion for maneuvers, such as midcourse navigation and maneuvers in the focal region to collect the image pixels. A JPL study (Liewer et al. 2000; Mewaldt & Liewer 2000) cited an Advanced Radioisotope Power System delivering 106 W weighing 8.5 kg (~12.5 W/kg). A system this small would be insufficient for boosting spacecraft velocity but might provide enough propulsion for small maneuvers and attitude control. Quantitative studies need to be done in a system design. The REP might boost the velocity by as much as 20%, e.g. 5 AU/year, albeit likely with a heavier system.

When we reach the focal region of the SGL, we must continue to fly along the FL for a flight time as long as it took us to get there, i.e., another 25 years. Images of the exoplanet would have to be constructed through a complicated deconvolution process of pixels sampled around the FL (see Figure 8). That is, the spacecraft would have to sample the image of an exo-Earth within a cylindrical region of a diameter of 1.3 km around the FL, while travelling at speeds of ~25 AU/year.

The major advantage of the solar sail spacecraft concept is that it offers a high solar system escape velocity and, thus, fast transit time to the SGL region. It also enables an interesting tradeoff for using a swarm of small spacecraft all using common external coronagraph (i.e., starshade). This could be the most desirable architecture for the SGL imaging mission architecture. These spacecraft could use smaller on-board telescopes (~50 cm) and rely on a common external coronagraph placed on a separate spacecraft and used to block solar light. Clearly, the formation flying aspect of operating the swarm is important for this architecture. It will be addressed in our Phase II study.

### *3.3.3 A string-of-pearls*

Another possible architecture that we addressed during the Phase I study was the "string-of-pearls", a set of small spacecraft launched together every 2–3 years. Advantages of this architecture include its inherent redundancy, which is important for very long duration flights. This is achieved by having ~5 of sensors in each "pearl", and perhaps 5–10 pearls flying in ~1 year intervals along the trajectory towards and in the vicinity of the focal line.

We deliberately chose such an open, flexible, and evolving architecture as it allows us to explore the boundaries of the entire mission tradeoff envelope. By loosely constraining the mission architecture with a common repeating component—the pearl—we were able to study a new class of missions that today would be discarded based solely on the perceived high cost. However, given the rapid development of the launch and spacecraft technologies, the cost and complexities of the missions today may have nothing in common with what we will be flying in, say 20–30 years. This approach allowed us to explore the feasibility of the mission and spacecraft requirements and otherwise explore the tradeoff envelope for missions capable of operating at >550 AU while collecting the imaging and spectroscopic data and also communicating back to Earth.

The technology requirements (e.g., reliability: continuous operation for > 50 years, system adaptability, and on-mission learning) were of particular interest. As many spacecraft systems are already being designed in anticipation of radical changes, focusing on resiliency, adaptability and disaggregation, we expect major progress in these areas in the near future, benefiting a potential SGL mission.

#### *3.3.3.1 Analysis of tradeoffs*

We approached this part of the study with the topics revolving around autonomy, positioning, navigation, communication, and on-board processing, focusing on the mission architecture and application of small spacecraft. This includes operations in the vicinity of the focal line while imaging the Einstein ring over a period of years as the spacecraft flies along the focal line, possibly with cooperative small spacecraft at multiple locations.

There are many elements to the SGL mission. For Phase I of the effort, we chose to study those that, from the perspective of mission requirements, come from the anticipated knowledge regarding position, navigation and timing (PNT).

Given the mission objectives, our initial tradeoff was to compare the spacecraft options as the mission could be accomplished either by a single spacecraft approach or via multiple spacecraft launched periodically in a string-of-pearls configuration (Figure 26). The logic behind those arguments is as follows (applied to all mission concepts considered):

- A single spacecraft that must operate for to 40+ years (~30+ years to reach the SGL region and for 10+ more years to take the data) has a limited probability of surviving, especially on the first attempt. Although Voyager 1 and 2 as well as Pioneer 10 and 11 operated at large heliocentric distances, they were not designed to do that as their primary missions. Although we learned how to operate these spacecraft at such large distances, these achievements cannot be taken as the baseline rule for a deep space flight.
- A single spacecraft would have to be programmed in advance to collect the needed data, in an environment in which the nature of the needed measurements and the conflicting optical signals are poorly known. To wit, it would have to study the environment and develop an evolutionary approach to survive in such a long flight. It would have to be able to adapt to different operating protocols characteristic for the transit phase (with very few critical events) to the operational phase (with active maneuvering, communication and sensing).

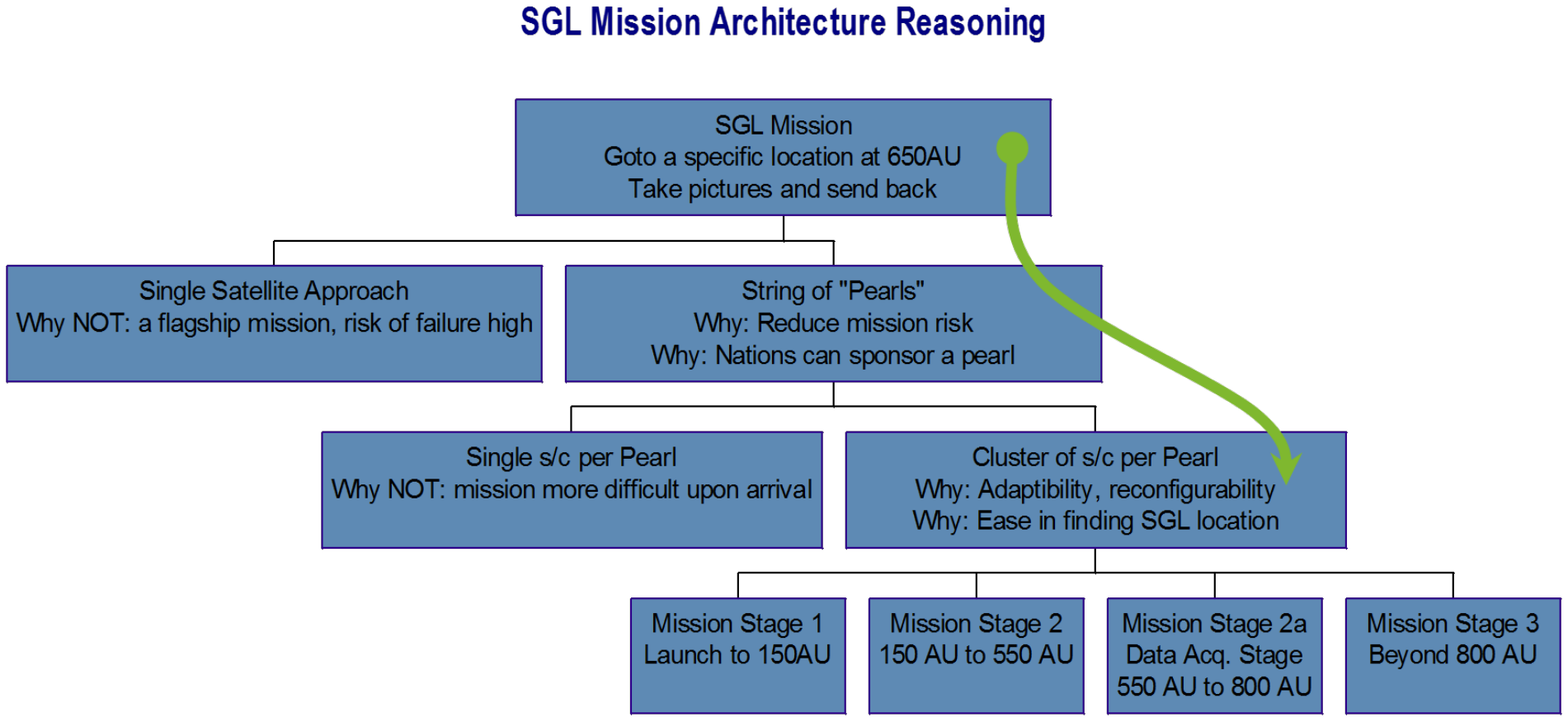

Figure 26. The tradeoff tree illustrates the advantages of the multi-spacecraft multi-launch "string-of-pearls" approach to the overall SGL campaign of missions.

- Conversely, a mission utilizing a number of cooperating spacecraft (string-of-pearls) with failure correction protocols would have a much higher probability of survival. With the string-of-pearls configuration, the arrival of the first pearl and the data it collects can be transferred to the next in line, enabling mission adaptability and machine learning. As the first of the pearls traverses the focal line, much will be learned about the optimum methods of collecting good data and rejecting noise (such as solar corona effects or errors in navigation and guidance). Given the significant unknowns in this process, learning by experience and adaptation of the operational protocols will be crucial to the collection of planetary images.
- The final set of images taken by the pearls architecture is the collection of millions of frames (snapshots), taken by many spacecraft within each pearl and along the entire string over several decades. The string-of-pearls configuration enables the use of optical communications by establishing relay stations en route for efficient, low power return of the image data.

#### *3.3.3.2 Mission architecture details*

To explore the string-of-pearls mission architecture, we postulate that each pearl is a cluster of spacecraft flying in formation. Figure 27 gives a notional view of 9 spacecraft making up a cluster that forms a single pearl. In this example, the formation has a single spacecraft in the middle and several in the periphery. The center spacecraft within each pearl helps to define the flight path vector for the string-of-pearls, aligning the communications link and ensuring that the attitude of the cluster is normal to the flight path vector (i.e., it coordinates the attitude and spacing of the peripherally stationed spacecraft). The peripheral spacecraft play an important role upon reaching the mission location.

As seen from the focal line of the SGL, the image of an exoplanet is within a cylinder with a diameter of ~1.3 km. For a given heliocentric position, finding this ~1.3 km cylinder within a radius of 10–50 km might pose a challenge. However, using a formation flying concept, it is possible to establish a concept of operations (CONOPS) that sends 2–4 spacecraft in a raster motion with increasing radii in search of the "tube" while others in the periphery wait for feedback.

The cluster concept also mitigates risk in the event of the failure of a single spacecraft in the pearl. The tradeoff study currently envisions ~20 annual pearl launches (~30 AU apart). The first pearl reaches the focal region of the SGL in ~25 years, and flies along the focal line for another ~ 8 years – the final pearl culminates the mission data collection 20 years later. Further advantages of this approach are that it requires a spacecraft design that is mass producible. It is an open architecture design in which any space-faring organization can contribute spacecraft or components while adapting and learning from previous pearl launches, and spreads the budgetary impact over several decades while capitalizing on any technological advances which may emerge over that time.

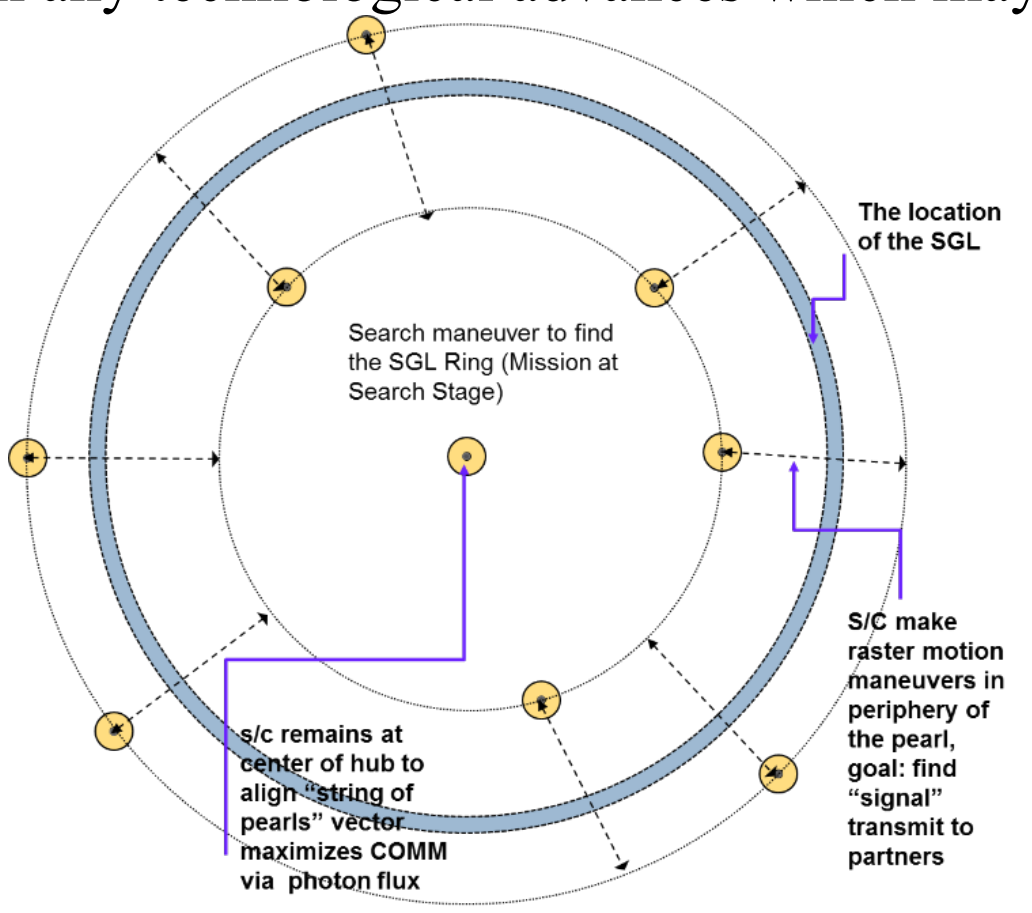

Figure 27. Cluster formation in a single "pearl" within the string-of-pearls configuration as observed in the trajectory path in the SGL's focal area.

Given the overall architecture, we looked at possible solutions for PNT. One can break down the mission into four stages, which may need different approaches to solve the PNT issue:

- From launch to the operational limit of the NASA Deep Space Network (DSN), which, in its current configuration (and near-future upgrades) may be able to operate out to ~200 AU;
- Cruise phase to the entrance into the focal region of the SGL (from 250 AU to 550 AU);
- Finding and tracking the focal line of the SGL (~550-600 AU); and
- data acquisition, keeping the spacecraft aligned with the focal line (beyond 650 AU).

In this scenario, the DSN would provide communication and navigation up to ~200–250 AU, but the diffraction-limited performance of the RF signal would limit its utility at ~650 AU. The handoff error from the DSN to the cruise phase would be of the order of 10 meters, such that the speed error will be small enough to allow the transition of the cruise phase by "dead reckoning" as the range error is not critical to the end game capture. The azimuth and elevation are critical and would be measured astrometrically by the onboard telescope (already available on board to support the primary mission objective), with each measurement repeated millions of times en-route to drive down the error to the needed levels. Rare course corrections (perhaps once every 5 years) would keep the pearls properly aimed.

#### *3.3.3.3 Clock synchronization*

Synchronizing the timing of the clocks on board is an interesting issue. One approach is to create a "pearl time" (independent of Earth time) in which all the spacecraft in a pearl are synchronized only with each other. This allows them to pool time-tagged data for transmittal down the string to Earth. A further step would be to synchronize the time of each of the string of pearls with each other. A final step would be to synchronize the pearls with Earth time.

For synchronization with Earth time, a chip scale atomic clock (CSAC) at the current technology produces a timing error of roughly 100 μs over nine days. Replacing the CSAC with a current grade GPS atomic clock reduces this error to 0.15 μs but at the cost of increased power consump-tion and weight. Regardless, no clock can maintain synchronization over the duration of the mission. By implementing an optical communication link with repeaters to Earth, elements of both P and T could be solved.

The optical telescope onboard could be used to support the communication link. A 100 W pulsed laser (e.g. 1 MHz repetition rate) on Earth at a 5 m telescope (similar to that at the Palomar Observatory) where the repetition rate is synchronized with a clock standard could serve as a reference to provide synchronization of all the clocks out to 650 AU. Moreover, by using simple on-off key (OOK) encoding, UTC data could also be sent if necessary. The end sensor could be a Geiger-mode avalanche photo diode. This approach also provides a means to measure range.

An alternative way would be to implement optical navigation that relies on optical ranging and astrometry, to be addressed in the Phase II study.

#### *3.3.3.4 Communications*

Lower power ~5W and low duty cycle (~1%), lasers on each pearl would be used for the communication links (up/down) along with the concept of cluster and formation flying (as depicted in Figure 27). They may also allow for alignment of the vector comprising the string-of-pearls. This becomes possible because of the low divergence requirements necessary for the communication laser and the optical spatial distribution of a high-quality laser. At an inter-pearl range of 30 AU (e.g. next repeater station), the photon flux distribution transverse to the direction of travel (i.e., in the plane of the formation-flying cluster) should mimic the laser spatial distribution, which is nominally Gaussian. The center spacecraft in the cluster could periodically maneuver (do cross-path motions) and peak up the signal by just peaking the photon flux. Once this is done, it would send commands to the peripheral spacecraft to reorient the formation.

Star trackers using both the Sun and numerous stars acquire data to maintain the plane comprising the cluster normal to the flight path. It is also possible to use the location of the planets within our solar system to get angular measurements referenced to the Sun. The CONOPS for the flight duration entails periodic time synchronization updates from earth, slight modifications in the flight vector via cross path motion coupled with long integration exposures using the star trackers to get high angular resolution data. The mission pushes angular resolution limits to fractions of arc seconds. However, during the journey to the SGL time is an asset. We suggest the use of efficient electric thrusters for the small, but necessary, trajectory changes. The $\Delta V$ fuel requirements are bound to be very small if the spacecraft mass can be kept small, ~10–20 kg.

#### *3.3.3.5 PNT CONOPS at mission location*

The SGL contour of the target planet is a conical shape going from the FL outwards. We assume that the observational period of the candidate planet is only ~40% of the viewing time, so that the CONOPS can be analyzed for the case of examination of several planets in a single exoplanetary system.

The arrival to the SGL is predicted by extrapolation of the DSN hand-off data. Attitude is determined from star tracker data supported by the optical sensor on the center spacecraft. The use of the data collection sensor as the feedback sensor for the navigation process is essential. Response to navigation commands within the cylindrical SGL focal region is via electric propulsion. As described above, we maintain the formation pattern, except that upon arrival, a search algorithm is initiated that puts into transverse motion the peripheral spacecraft keeping the center spacecraft (i.e., the coordinator) fixed.

The concept entails the peripheral spacecraft to make, with ever increasing radii, raster type motions using a combination of sensors to advantage. The peripheral spacecraft all have telescopes because members of a cluster are interchangeable. The search routine entails that the spacecraft acquire (single-pixel) data at high speed with timestamp.

1) Using a visible light sensor, we anticipate that the luminosity of the solar corona (known spectroscopically with high fidelity) to decay as a function of the distance from the Sun (Figure 13).
2) We anticipate that when the parent star of the desired exoplanet comes within the focal region of the SGL, the luminosity at that visible light sensor should abruptly increase.
3) We use this increase in luminosity to then switch on a second sensor, which is fitted with an IR filter (1–5 μm) looking for molecular emission (e.g. water, methane). This should provide high contrast to the emission from the Sun and the parent star.
4) Signals from the second IR sensor are also timestamped.
5) The search ends when the signal from the parent star emission decays. The start and stop timestamps and the spacecraft's velocity (measured relative to the spacecraft in the cluster center) are transferred to the awaiting peripheral spacecraft which modify their transverse velocities and start their search.

The first set of spacecraft begin the return traverse following the same process. The goal is to alter or adapt the cluster formation such that the peripheral spacecraft make periodic traverses of ever shorter range through the ring. Data are acquired and transmitted to the center spacecraft when in the ring. The center spacecraft communicates all information to the approaching pearl, which passes along the information but also makes adjustments to its own formation and approach. A nine-spacecraft cluster would allow the sending of 4 spacecraft in the cardinal directions of the cluster coordinate system, followed by the second 4.

Finally, it is also expected that the location of the SGL ring and the sensor signal pattern (visible, IR etc.) will change as the first pearl flies along the 200 AU data collection tube. Data about these changes will continually be fed to the next pearl and to subsequent approaching pearls, such that the overall system (i.e., the string-of-pearls) learns and achieves improved performance.

#### *3.3.3.6 ΔV for PNT*

We conducted a first-order analysis of the propellant usage and ΔV requirements for navigation and position maintenance. The primary adjustments are 1) tracking corrections in the outbound or cruising portion, 2) phasing corrections during the SGL search and exoplanet capture phase, and 3) the fact that the exoplanet is orbiting a parent star and will be moving during the imaging phase.

Modeling after Earth's orbit around the Sun, a spacecraft in the pearl would need ~30 m/s of ΔV per year to keep an exoplanet within the focal zone. Assuming a 20-year mapping mission (traversing the SGL data collection tube) yields a ΔV of 600 m/s. The ΔV needed for trajectory correction for the outbound, cruise portion depends on two factors – (1) the initial correction needed due to the injection error (the difference between the desired and the actual state vector at release of the s/c) , and (2) once the injection error is corrected, the further accumulated errors due to drift of the s/c trajectory from the desired flight path. For the first of these, the DSN tracking errors are negligible – so the ΔV is only driven by injection error. This needs to be quantified in Phase II but for now we postulate that 400 m/sec should be sufficient to correct the injection error.

For the second set of corrections (during the 20-year cruise to the SGL) we can assume that the handoff from the DSN (at perhaps 200 AU) contributes a negligible error in range (range error is unimportant as we are aiming at a semi-infinite line), but the cross-track error must be adjusted by on-board propulsion. A DSN cross-track velocity error of 1 cm/sec at 200 AU corresponds to a cross-track position uncertainty of $10^4$ km CEP at arrival at the FP.

We can measure such an error by multiple on-board measurements of the angle between the sun and a few bright stars, and make small corrections on the way (perhaps a correction every 5 years). We allocate 75 m/sec would bring the spacecraft to the FP. This puts the outbound journey ΔV requirements at 475 m/sec. The resulting total ΔV for the mission then becomes 1,075 m/s. We then consider a total ΔV requirement, including a margin of error, of 1,200 m/s. If a propulsion system with 8,000 s $I_{sp}$ thrusters are used, the propellant mass fraction is only 3% (i.e., 3% of the initial spacecraft mass departing Earth will be propellant). The stressing requirement is not the propellant weight, but the energy needed, which for a 100 kg s/c corresponds to some ~600 kW-hrs (at 0.05 N/kW), another reason to get the weight down to the 10–20 kg level.

The analysis shows that the highest trajectory adjustment occurs during the outbound phase where corrections on the order of 95 m/s are needed over a 5-year timeframe. One efficient thrusting approach is to produce the ΔV over a period of 1 year or less for each 5-year trajectory-update. Providing 95 m/s over 1 year requires a thrust of only 3 μN per kg of spacecraft mass. Power and thrust are proportional to spacecraft mass. During the data acquisition phase, the yearly ΔV requirement drops to 20 m/s per year.

# 4 SUMMARY AND RECOMMENDATIONS

The prospect of getting an image of an exoplanet and to spectroscopically detect and characterize life being there is compelling. New coronagraph design, even if it is never used at SGL distances, would benefit the exoplanet search community. The mission strongly suggests the development an architecture that relies heavily on autonomous action, adaptability and the ability to “learn”. These capabilities would benefit Mars missions or other missions to the outer planets where communications are delayed. If this mission provided the indisputable spectroscopic proof of life on an exoplanet, then it would provide one of the most provocative pieces of scientific discovery ever!

The results of this investigation could be catalytic. Our approach is radically different compared to existing ones arguing for instruments with ever larger aperture size, e.g. HST to JWST to LUVOIR to other 15-30 m diameter large space telescopes (ATLAST++). Because of its broad nature, any technology developed in support of the SGL mission will benefit the astronomical community. Our study could lead to novel designs of coronagraphs, high-precision navigation, autonomous, intelligent and adaptive systems, instruments designed to operate for decades. All these technologies are essential to other NASA missions. However, perhaps the more compelling benefit is how to implement missions to 500-1,000 AU over the next 40 years with the benefit of enabling investigations throughout our solar system, the Kuiper Belt region, and to the Oort cloud.

## *4.1 Summary of Phase I results*

Detection of signs of habitability via high-resolution imaging and spectroscopy of an exoplanet is the most exciting objective of a mission concept to the focal area of the SGL. The work during Phase I was directed at the development of the instrumentation and mission requirements, and also to study a representative set of mission architectures.

First of all, to provide a solid understanding for the optical processes of image formation, prior and in parallel to the main effort, we developed a wave-theoretical treatment of the SGL. This treatment accounts for the spherically symmetric gravitational field of the Sun which was treated in the harmonic gauge at the first post-Newtonian approximation of the general theory of relativity (Turyshev 2017; Turyshev & Toth 2017). The approach allowed us to study all important optical properties of the SGL needed to design an astronomical telescope, i.e., to describe the PSF, resolution, magnification, plate scale, etc.

As an extension to this work, we improved the model for the SGL by including the effects of the solar corona (Turyshev 2018a) and that of the gravitational multipole moments of the Sun (Turyshev 2018b) on the light propagation in the immediate solar vicinity. This work in particular allowed us to have a more realistic description of the SGL, which accounts for the solar corona, solar oblateness and solar rotation. The improved PSF is yet to be used for simulation and related mission design, but it already provided us with very important insight – the most important contribution to the SGL is that from the monopole, while all other effects are small and well modeled.

This work allowed us to develop a comprehensive understanding of the image formation by the SGL and the technology needs for a realistic mission, data collection and image deconvolution. It was guiding us in the instrument and missions design work conducted during Phase I of the effort.

Specifically, during the Phase I we were able to accomplish the following major tasks:

<u>Task I: Development of the system and mission requirements</u> to guide the preliminary design concepts and formulate key mission, system, and technology operation requirements:

- SGL's optical properties (Turyshev & Toth 2017) led to a solar coronagraph design capable of blocking the solar light to the level of the solar corona at a given position of the Einstein ring. The design resulted in a coronagraph with $2\times10^{-7}$ suppression, meeting the requirements. The 10% coronagraphic throughput yielded a 2-m telescope. We identified the instrument/mission design parameters which could reduce the telescope's size, namely i) a more advanced occulter mask, an external starshade solar coronagraph, iii) operating at larger heliocentric distances.
- To demonstrate imaging with the SGL, we investigated the application of rotational deconvolution and have shown that the SGL allows for a 300 ×300 pixel image of an exoplanet. We also estimated the effectiveness and integration time for a direct deconvolution. Our estimates show that a 500×500 pixel image of an exoplanet is possible with ~2 years of integration time by a direct deconvolution approach, suggesting exciting solutions to the imaging problem.
- We addressed the question of finding and studying life on an exoplanet based on the set of primary instrument observables. With a very respectable spectroscopic SNR $10^3$ in 1 sec, we conclude that the signal will be sensitive to disturbances in the atmosphere of an exo-Earth; it will be able to detect methane, oxygen and likely other molecules. An added benefit is that the same mission may also be able to take images of all the planets orbiting that star.
- We identified a key mission design driver that was not previously considered in an SGL-related studies. This driver is the reflex motion of the Sun with respect to the solar system BCRS, primarily due to gravitational pull from the giant planets, Jupiter and Saturn. That motion is slow and predictable, but it requires that the spacecraft have sufficient propulsion capabilities on board. We observed that the reflex motion induces smooth and easy implementable changes in the cross-track trajectory of the spacecraft, thus, reducing the $\Delta V$ requirements.

<u>Task II: Identification and study of possible mission architectures:</u> Initially keeping the design envelope wide to allow assessment of key mission, system, and technology operation drivers:

- We formulated the requirements for a mission that could deliver a healthy and capable spacecraft to heliocentric distances beyond 700 AU, place it on an actively controlled trajectory, and form a telescope that could exploit the unique optical properties of the SGL.
- We considered several mission concepts involving a single spacecraft, a small spacecraft with solar sails, and a cluster of mid-size spacecraft. Our results indicate that an SGL mission to for direct imaging and spectroscopy of an exoplanet is challenging, but feasible. We identified the design trade parameters that could lead to a robust mission and improve its performance.

- We explored an architecture that relies on a pair of spacecraft connected by a boom (or tether) of variable length (DeLuca, 2017). We also explored the role of spacecraft swarms or clusters to reduce the navigational/maneuver requirements to capture images at higher speeds. We considered the effect of the proper motion of the exoplanet, its orbital motion, and rotation on the imaging satellite requirements from the perspective of the necessary $\Delta V$.
- We investigated CONOPS of a spacecraft at SGL for detecting, tracking, and studying the brightness of the Einstein ring around the Sun. Our baseline approach relies on optical comm/nav, utilizing lasers and precision optical astrometry. We considered a set of instruments and on-board capabilities needed for unambiguous detection/study of life on another planet.

As a result of these efforts during Phase I, we were able to demonstrate the feasibility of a mission to the SGL for the purposes of studying life on an exoplanet. We have identified the next steps needed to improve our understating of the entire mission design envelope as it relates to moving spacecraft around the SGL's instantaneous focal line.

### *4.2* Towards a realistic SGL mission concept

To build a $10^3 \times 10^3$-pixel image, we need to sample it in a pixel-by-pixel fashion, while moving with resolution of ~1 m. This can be achieved relying on a combination of inertial navigation and laser beacons s/c placed in 1 AU solar orbit whose orbital plane is co-planar to the image plane.

Tethering or (nuclear-)electric propulsion could be used to perform raster scanning with a spacecraft >550 AU away (see DeLuca, 2017). One way to scan the image of an exo-Earth is to conduct a spiral scan to follow the planetary motion while using a ~1.3-km tether and the RTG on the other end of it (to balance the s/c). This reduces the fuel requirement for raster scanning the image.

A mission to the SGL is challenging, but not impossible. Given the current state-of-the-art, several technologies may enable a meaningful step beyond our solar system to distances of 550–1,000 AU in 25–35 yrs (Stone et al. 2015; Alkali et al 2017). We will consider precision navigation of an SGL spacecraft in the image plane and will investigate the use of laser beacons in a 1 AU heliocentric orbit for communicants, guidance and navigation.

We will consider a long-term technology development program with the following ten topics: i) Mission and trajectory design needed to achieve high escape velocity and shorter flight time, but also with small orbit injection errors; ii) Propulsion systems, such as nuclear-thermal, nuclear fusion, nuclear fission, solar thermal propulsion, laser-beamed energy, laser ablation, solar sails, electric sails and more; iii) Power systems, including nuclear power; v) Structures, such as lightweight multifunctional structures, deployable structures, etc. vi) Thermal design and stability, low-power, low-temperature systems, etc. vi) Telecommunications systems utilizing both RF and optical communications; viii) GNC, including spacecraft stability, pointing to Earth. ix) Avionics systems to support long term survivability and autonomous operations; and x) Instruments and payload, including highly miniaturized solutions.

### *4.3 Topics for further study*

Our results of the Phase I study demonstrate the feasibility and the challenges for a mission dedicated to imaging with the SGL. This earlier work provides us with a solid foundation for the next phase of this exciting effort. Therefore, during Phase II, we plan to continue to explore the topics that enable a robust SGL mission, including refinement of the mission architectures discussed here by taking them through mission simulations and design trades.

In addition, we would like to consider the following seven major topics:

1. To investigate the science operations in support of the primary objectives: high-resolution imaging and spectroscopy. We will explore the ways of detecting photons from the Einstein ring, collecting them in a 4-dimensional data cube, processing and deconvolution. Insights on the image formation will improve the mission concepts, yielding realistic mission requirements.
2. To study of direct and rotational image deconvolution approaches in a realistic setting including effects of planetary rotation, varying planetary features (i.e., time-variable clouds, seasons), telescope pointing errors, etc. We would like to combine these approaches in a more generalized deconvolution simulation. We will look at the non-uniformity of light distribution within the Einstein ring, which may yield additional information for the deconvolution process.
3. To study an instrument comprising a swarm of small spacecraft, perhaps even launched together but each moving at a slightly different trajectory but parallel to the instantaneous optical axis. Such an instrument would rely on the light collection capabilities enabled by a formation flying architecture, taking full advantage of the SGL amplification and differential motions.
4. Given the enormous amplification of the SGL, to study the possibility of spectroscopy of the exoplanet, even spectro-polarimetry. It will not just be an image, but potentially a spectrally resolved image over a broad range of wavelengths, providing a powerful diagnostic for the atmosphere, surface material characterization, and biological processes on an exo-Earth.
5. To investigate the issue of imaging from the time-variability of the SGL system resulting from the solar motion with respect to the BCRS. To investigate how the s/c will raster an exo-planet as it optical axis moving on a $10^4$ km orbit. To study the relevant station keeping aspects.
6. To study mission design: i) flight system and science requirements; ii) key mission, system, operations concepts, and technology drivers; iii) description of mission and small craft concepts to reach and operate at the SGL; and iv) study instruments and systems for the SGL spacecraft, including power, comm, navigation, propulsion, pointing, and coronagraph. To conduct mission architecture trade studies aiming at PNT requirements for the SGL mission.
7. To conduct trade studies with a set of key driving parameters: a) heliocentric distance, b) telescope's aperture, c) integration time, d) detector type and sensitivity, e) coronograph/starshade performance, etc. Consider tradeoffs between a single telescope vs microsat system. A small telescope has limited capabilities, but opens up the possibility of sending multiple spacecraft.

The primary innovation in this proposal comes from the utility of using a well-positioned sensor at the SGL to image and raster scan a promising exoplanet. Other innovations include: i) the recognition and intentional use of the SGL to magnify with unprecedented angular resolution, ii) the CONOPS of acquiring the information pixel-by-pixel with calibrated and time synchronous s/c motion, iii) deconvoluting the image using the known PSF. Given the state of current astronomical telescope developments, including JWST and TESS, our proposal is timely and relevant. It takes advantage of the likely discovery of numerous exoplanet candidates by TESS and further assessed by JWST. While these missions may provide "hints", one image from the SGL could "seal-the-deal". The SGL mission offers a unique means for determining exoplanetary atmospheric chemical composition and defining the propensity for habitability of life.

To date, all results look promising, both for getting there and for capturing high resolution images with spectral content. Technological considerations with regards to mission architecture, instruments, comm etc. also look feasible. The mission has the potential of being the most (and perhaps only) practical and cost-effective way of obtaining kilometer scale resolution of an exoplanet.

Concluding, we emphasize that it remains to be determined just how complex the capture and creation of direct images of an exoplanet will be using the SGL. The cost of a mission to the SGL focal region also remains to be determined. However, if it does prove to be a feasible mission, there may be cost and science tradeoffs between remote sensing using the solar gravity lens and flying to, operating and returning data from a planet in another star system many light years away. In any case, the first job is to simulate image creation in the SGL (Turyshev & Friedman, 2017). Although we investigate the question of spacecraft design and reaching the extremely distant regions outside the solar system, emphasis is placed on the feasibility of mission operations in support of the primary science objectives: high-resolution imaging and spectroscopy.

***Acknowledgments:***

We would like to express our gratitude to our many colleagues who have either collaborated with us on this manuscript or given us their wisdom. We specifically thank Louis D. Friedman who provided us with very valuable comments, encouragement, support and stimulating discussions while this document was in preparation.

The work described here was carried out at the Jet Propulsion Laboratory, California Institute of Technology, under a contract with the National Aeronautics and Space Administration.